# Stirring Astronomy into Theology:
## Sir Isaac Newton on the Date of the Passion of Christ

### Ari Belenkiy
Mathematics Department, Bar-Ilan University, Ramat Gan 52900 ISRAEL

and

### Eduardo Vila Echagüe
IBM-Chile, Providencia, Santiago, CHILE

**Abstract.** It is known that Sir Isaac Newton suggested a date for the Passion of Christ in the posthumously published *Observations upon the Prophecies of Daniel and the Apocalypse of St. John* (1733). What was not known is that the first attempts to find that date were made during the early period of his life. The Jewish National and University Library in Jerusalem contains two undated drafts in Latin under the same title, *Rules for the Determination of Easter*, grouped as Yahuda *MS* 24E.

The earlier draft contains multiple references to the virtually forgotten *De Annis Christi (*1649*)*, written by Villum Lange, the 17[th] century Danish astronomer and theologian, who might have been Newton's first mentor on the Jewish calendar tradition. Lange educated himself in this subject from Maimonides and a certain Elia, whom we identified with Eliyahu Bashyatzi, a Karaite of the 15[th] century. The second draft shows not only Newton's close acquaintance with Maimonides' theory of first lunar visibility, but also his attempt to simplify the latter's criteria by introducing different, more practical parameters. These "astronomical exercises," announced in a 1673 book, were likely intended to appear as an appendix to Nicholas Mercator's 1676 book.

Both of Yahuda 24E's drafts contain an astronomical table with the solar and lunar positions for years 30-37, which Newton used to decide on the year and date of the Passion. The astronomical data comes from either 1651 *Harmonicon Coeleste* or 1669 *Astronomia Britannica* by Vincent Wing, a semi-forgotten astronomer of the seventeenth century. This makes Yahuda 24E one of the earliest of Newton's drafts, likely written in 1669-73 and certainly not later than 1683/4.

A comparison of the two drafts of Yahuda 24E shows that in the later one, Newton changed his allegiance from St. John's chronology of the Passion to that shown in the synoptic gospels. This mindset, as can be seen from his *Observations,* was dramatically reversed again in his later years, supported by a forced, somewhat unexpected interpretation of the Jewish calendar tradition.

# Introduction

In chapter XI, "Of the Times of the Birth and Passion of Christ," in his *Observations upon the Prophecies of Daniel and the Apocalypse of St. John*, written toward the end of his life and published posthumously in 1733, Sir Isaac Newton suggested the following explanation for his finding that the year of the Passion was year 34:

> Thus there remain only the years 33 and 34 to be considered; and the year 33 I exclude by this argument: In the Passover two years before the Passion, when Christ went thro' the corn, and his disciples pluckt the ears, and rubbed them with their hands to eat; this ripeness of the corn shews that the Passover then fell late: and so did the Passover A.C. 32, April 14 but the Passover A.C. 31, March 27 fell very early. It was not therefore two years after the year 31, but two years after 32 that Christ suffered. Thus all the characters of the Passion agree with the year 34; and that is the only year to which they all agree.

Here Frank Manuel somewhat ironically remarks that

> "It would not have been sensible for the apostles to eat unripe corn." [1]

True, Newton's reliance on Jesus' common sense in practical matters is unmistakable. He might also have realized a nuance that had escaped Manuel: ripe grain must be rubbed before eating to remove the husk, and the rubbing (threshing) – not just eating – was not allowed by the rabbis on Jewish holy days. The deeper question that Manuel neglected to ask concerns Newton's comment that "all the characters of the Passion agree," whereas we see only one!

William Whiston, Newton's notorious disciple, successor to the Lucasian chair and rival in religious matters, was in favor of year 33. In his *Remarks on Sir Isaac Newton's Observations upon the Prophecies of Daniel and the Apocalypse,* published in London a year after Newton's book, Whiston wondered:

> But why so great a Chronologer and Astronomer as Sir Isaac Newton, should not, in this Age, chuse rather to determine both the Birth and Death of Christ by proper Characters derived from those sciences, as all learned Men have endeavoured to do, than from some more uncertain Supposals or Suspicions of his own, I cannot imagine.[2]

---

[1] F. Manuel. *Isaac Newton: Historian* (Cambridge, Mass. 1963), p. 291.

[2] In: W. Whiston, *Six Dissertations* (London 1734), p. 308.

The manuscript, which was sold at Sotheby's 1936 auction and presently bears the name Yahuda MS 24E (Appendix 1), witnesses that Newton relied not only on "supposals and suspicions" but supported year 34 with solid astronomical computations.

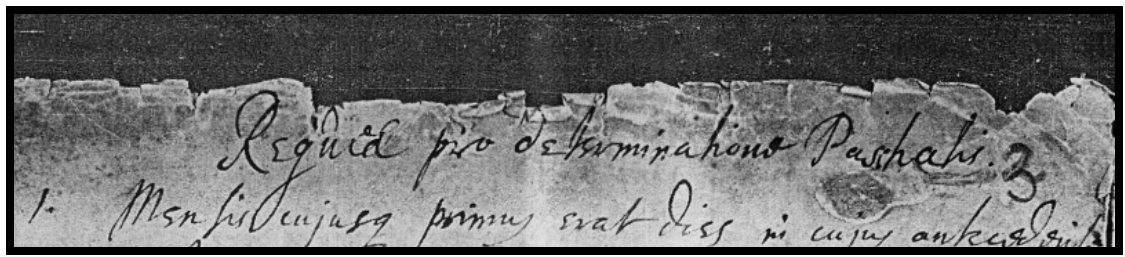

**Figure 1: Manuscript Title (Fair Copy) (Courtesy of JNUL)**

## Yahuda 24E: Two Drafts

The Yahuda 24E contains two drafts, both in Latin.[3] Both pieces have the same heading: *Regulae pro determinatione Paschae,* or *Rules for the Determination of Easter.* Their goal is to determine the date of Christ's crucifixion. Both start with a set of general rules, mainly for determining lunar visibility on a specific night. A table with astronomical data containing possible Easter dates for the years 30-37 A[nno] C[hristi] follows. Finally, each manuscript presents a unique (albeit conflicting) solution for the year of the Passion.

Newton scholar David Castillejo, who sorted all of the Yahuda MSS in 1969, called a two-page draft the 'rough draft' and another three-page draft the 'fair copy.' The 'rough draft' is written in Newton's hand. The 'fair copy' (see Figs. 1 and 4) is written in a hand different than Newton's, which is not surprising: several of Newton's theology manuscripts were nicely rewritten in the hand of his Trinity College roommate during 1663-83, John Wickins, while the entire fair copy of *Principia* was copied by Humphrey Newton, a roommate during 1683-88. The handwriting of the 'fair copy' shows compelling similarities with Yahuda MS 23, also attributed to Wickins.[4] For the sake of brevity, we

---

[3] The drafts are nearly illegible. Many characters have unusual shapes, are poorly visible, or are simply missing. The second author, Eduardo Vila Echagüe, used all his skills in Calligraphy, Latin, Astronomy, and Theology to extract the correct, or at least the more likely, reading. The language, Latin, and the style of Yahuda 24E sharply distinguish it from the rest of Yahuda 24 dealing with Newton's unpublished calendar; see A. Belenkiy and E. Vila Echagüe, "History of One Defeat: Reform of the Julian Calendar as Envisaged by Isaac Newton," in Notes & Records of the Royal Society of London, 59 (3), September 2005, pp. 223-254.

[4] See R.S. Westfall. *Never at Rest. A Biography of Isaac Newton* (Cambridge Univ. Press 1980), 319, n. 111. At our request, John T. Young of The Newton Project < http://www.newtonproject.sussex.ac.uk > confirmed this identification, comparing the handwriting with a printout of the fair copy of the *Hypothesis Concerning*

assign to the 'fair copy' the name W (Wickins), and to the 'rough draft' the name N (Newton).

Though N carries many amendments that at times make the reading impossible, we will show that it was written later than W. Three circumstances can account for this fact: emendations of the text, rearrangement of the data, and style of references. The first is the most decisive: in W the dates of the Last Supper were first computed under an assumption that it took place on the eve of Nisan 14, though later this was amended by superscripts. These superscripts became the major dates used in N.

The table in W includes the years 30-37 AC. After postulating the relevant rules of the Jewish calendar and a brief discussion, Newton discards all of the years except 33 AC, because only in that year did the Last Supper (eve of the Passion), which he places on the eve of Nisan 14, fall on Thursday, April 2. He concludes that the Passion took place the next day, on Friday, April 3, 33 AC.[5] In the end, he alerts the reader to the possibility that the year of the Passion could have had *intercalated*, i.e., augmented with an additional, 13th month. Only year 34 has to be checked, but Newton discards this option as well (because the eve of Nisan 14 falls on Wednesday) and confirms the initial choice of year 33. Though in N the same table is included, in its last line the data for year 37 is replaced by the data for year 34 with the additional month, whereas the latter was presented separately in W. While N solely relies on Maimonides, W relies on a virtually unknown Danish theologian and astronomer, Wilhelmus Langius, also known as Villum Lange.[6]

## Lange's De Annis Christi

Lange's 1649 *De Annis Christi* comes as a complete surprise, fresh wind in old sails. No previous academic study on Newton, in particular on Newton's close association with the ideas of Maimonides, has ever mentioned the book, though it predates by 20 years de Veil's Paris translation of Maimonides' astronomical work.[7]

---

*Light and Colors*, CUL 3970.3, ff. 40-466, which is unquestionably written in Wickins' hand. The differences are most noticeable in the letter "e" and the upper parts of the numbers 3 and 5.

[5] This was the date suggested by J.J. Scaliger in the *second* edition of *De Emendatione Temporum* (Leiden 1598), pp. 525-6.

[6] Lange (1624-82) served as director of the Royal Observatory in the Rundetaarn in Copenhagen in 1652-82.

[7] Harrison's catalogue (*The Library of Isaac Newton*, Cambridge 1978) does not list Lange among Newton's personal books, but the Cambridge University library contains Moore's copy of Lange's book, dog-eared in Newton's fashion.

When asked for Newton's sources on Maimonides' astronomy, experts usually quote *Kalendarium Hebraicum* by Sebastian Münster and *De Sacrificiis Liber* by Ludovicus de Compiegne de Veil. Richard H. Popkin[8] and Matt Goldish[9] pointed out that the 1527 Basel edition of the *Kalendarium* and the 1683 London edition of de Veil's Latin translation of the *De Sacrificiis* (with *Majiemonidae Tractatus de consecratione calendarum et de ratione intercalandi* included) were in Newton's library. Jose Faur,[10] however, specifically referring to the Yahuda collection, suggested that Newton's quotations from Maimonides were taken directly from the 1669 (Paris) edition of de Veil's *Secunda Lex, Tractatus de conservatione calendarum* with the Latin translation of Maimonides' *Sanctification of the New Moon*, though the Cambridge University library carries not a single copy of it.[11] If Faur's conjecture is true, then the year 1669 would be a precise borderline between the two drafts, because then Newton could have begun reading Maimonides directly, not just gleaning separate quotes from Lange.

Speaking about the date of the Passion, Lange gave a review of Jewish astronomical lore, mixing Maimonides with Elia Scripturarius. Elia was identified as Eliyahu ben Moses Bashyatzi (d. 1490), the Karaite, whose book אדרת אליהו [Aderet Eliyahu] was published in Constantinople c. 1540.[12] Excerpts from the *Aderet Eliyahu* were made available by John Selden in 1644.[13] Elia's work summarized the theory of the first lunar visibility, developed by al-Battani (c. 858-929), Abraham Ibn Ezra (c. 1092-1167), Maimonides (1135-1204), and Emmanuel the Geographer (13th century), also providing new criteria and tables for first lunar visibility for the latitude of Constantinople, where Elia resided.[14]

Lange's goal was to disprove dates for the Passion given by Joseph Justus Scaliger as April 3, 33 AD, and by Dionysius Petavius as March 23, 31 AD.[15] Lange rightly rebuked both for their neglect of the lunar visibility problem, as the core of the Jewish calendar

---

tradition.[16] In turn, Lange pioneered the following primitive astronomical approach. On page 409, Lange provided a table somewhat similar to the one in Yahuda 24E, with the time of the *mean* lunisolar conjunctions in Jerusalem in two spring months of the years 31-35.

| Year | Mean Conjunction | | Roman & Jewish Weekday |
|---|---|---|---|
| XXXI | Mar 06:48:51 | 12 | II<br>II |
| | Apr 19:32: 55 | 10 | III<br>IV |
| XXXII | Mar 04:21:34 | 30 | I<br>I |
| | Apr 17:05:38 | 28 | II<br>II |
| XXXIII | Mar 13:10:14 | 19 | V<br>V |
| | Apr 01:54:18 | 18 | VII<br>VII |
| XXXIV | Mar 21:58:53 | 08 | II<br>III |
| | Apr 10:42:53 | 7 | IV<br>IV |
| XXXV | Mar 19:31:35 | 27 | I<br>II |
| | Apr 08:15:39 | 26 | III<br>III |

**Table 1. Lange's Table from *De Annis Christi*, p. 409.**

An interesting feature of Lange's table is that it assumes for each year two possible dates for Passover, implying an embolismic month – an idea picked up later by Newton. Lange's mean conjunctions were only 4 minutes and 57 seconds later than Jewish Molad timings.[17]

Lange's own theory on the date of the Passion was extravagant – he placed the <u>Crucifixion on Thursday</u>, relying "on the authority of the Holy Scripture" [pp. 414-5]. Thursday could be a solution for those preferring a resurrection "in three days" to the traditional "on the third day." Lange's resolution of the synoptic gospels vs. St. John problem was also new: <u>St John referred to a Passover observed by the Jewish authorities, while the synoptic gospels described the personal Passover of Jesus, which took place one day earlier</u>.[18]

---

[16] Incidentally, Scaliger's date agrees with the full moon, while Petavius' date is three days earlier than the full moon of Nisan, a Jewish lunar month when Passover occurs on the 14th day.

[17] The table suggests that Lange used a mean month of 29d 12h 44 min 3s and 17 thirds. However, the solar and lunar positions on page 128, which Lange says were obtained using Parisian tables, are not compatible with Lange's tables on pages 55-58, which imply a mean month of 29d 12 h 44min 3s and 12 thirds. Parisian tables were printed in Paris around the mid-XVI century, spuriously said to be based on the *Alphonsines*.

[18] Modern researchers are unaware of Lange's priority in this interpretation. See, for instance, discussion in "Dating the Crucifixion" by C.J. Humphreys and W.G. Waddington (Nature, 306, 1983).

Though the contemporary Passover could not fall on Friday,[19] Lange spent pages [pp. 396-9] quoting the Talmud (e.g., tractate *Menachot*) and later rabbinical authorities (Ibn Ezra, Isaac Arama, Elia of Paris) to show that the first day of Passover in practice historically often did fall on Friday. With that, this day was, according to Lange, a public (Pharisaic) Passover, the "Chagiga." Non-Pharisees, like the Sadducees and those who resided far from Jerusalem and did not want to be dependent on the central authority, could have celebrated Passover one day earlier, judging by personal observations of the young moon.

Lange [p. 405] says:

> We shall make this manifest using an example. Let 19 March be 30 days from the previous sighting, or day 30 of month Adar (as was the case in year 33 A.D., as we shall show later). Then in that day both the Sanhedrin (whose members were mainly followers of traditions) and the Scripturarii looked for the Moon, but could not see it. The true New Moon happened in Jerusalem after noon of 19 March. It was not possible on that day to see the Moon before night. Therefore Talmudists established that 20 March was the Neomenia [Rosh Chodesh] of the month Nisan.[20] But, according to the Karaites, the Moon could be seen in the land of Israel 25 hours after Syzigia [True Conjunction], as we said in book I.[21] Let 25 hours be added to the true New Moon of March (which in Jerusalem occurred 2:31 hours p.m. 19 March[22]). The result is 20 March 4 hours p.m. There were still more than 2 hours until night (the third 'vesper' of the Karaites). Then the day that preceded 20 March, that is 19 March, was the Neomenia of the month Nisan, according to the Karaites, though for the Talmudists it was 20 March.[23] Therefore 14 Nisan according to the Karaites fell on 1 April; but for Talmudists, 2 April.

In short, Lange computed that Molad Nisan fell after noon on March 19. Therefore, the moon was NOT visible that night. But Lange argued that on the next evening (March 20), the moon was large enough and likely seen two hours prior to sunset. Therefore, the Talmudists have viewed the outgoing day (March 20) as Neomenia (Rosh Chodesh =

---

[19] This is because Rosh Hashana follows the festive day of Passover (Nisan 15) by exactly 163 days and cannot fall on Sunday. For historical perspective of the postponements (in Hebrew: *Dekhiyot*) and, in particular, Dekhiyah A which forbids Rosh Hashana falling on Sunday, see S. Stern, *Calendar and Community*, Oxford Univ. Press, 2001, 191, or A. Belenkiy, "A unique feature of the Jewish calendar – Dekhiyot" (*Culture & Cosmos*, 2002, 6 (1), pp. 3-22).

[20] Rosh Chodesh could have taken place on March 20 automatically only if March 19 was the 30th day of the previous lunar month Adar – a fact Lange never verified.

[21] Actually 25½ hours, according to Elia [*Aderet Eliyahu*, ch. 29 = Selden, p. 65], whom Lange correctly quotes on p. 126.

[22] Lange does not provide details of the computation of the time of the true conjunction. According to his table above, the mean conjunction occurred at 1:10 p.m., March 19.

[23] Nor did Lange confirm whether the length of the previous month, computed by his method, allowed Talmudists to place Rosh Chodesh Nisan as early as March 20 (since any month cannot be less than 29 days).

Nisan 1), while sectarians could have placed the beginning of the month a day earlier. Apparently, Lange assumed that the Talmudists waited until very end of the 30th day to announce Rosh Chodesh,[24] while in case of the Scripturarii (Sadducees), whom he associated (wrongly) with the later Karaites,[25] he makes an exegesis on the brink of fault.

The Karaite Elia [*Aderet Eliyahu*, ch. 16 = Selden, pp. 61-62] had proposed a criterion of *two hours prior* to sunset on the 31st day (possibly claiming the first visibility was two nights earlier), and Lange [pp. 404-5] seemingly had understood him correctly. But in his own example, the moon on March 20 had to be just 13° above the sun two hours prior to sunset, and it hardly could have been seen at that moment.[26]

Assuming Jesus was associated with the Sadducees, he, according to Lange, could have held a private celebration of the Passover a day earlier than the day officially announced. Then the weekday of a conjunction that fell before 18:00 (March 19 = Nisan 1 according to Jesus) must be identical to the weekday of the public Nisan 14.

Lange's table allows conjunction (and hence Nisan 14) to be on Thursday only in year 33. Therefore, the date of the Passion must be April 2 (Thursday), 33 AD.

## Jewish Criteria for First Lunar Visibility

The data in W that Newton extracted from Lange [pp. 126-27] at first glance seem extremely disordered, like a remark that the Sanhedrin used to send appointed people to the tops of the hills to catch the first glimpse of the young moon; another remark is that even if a pair of witnesses arrived too late, their witness could be accepted *post factum*.[27] Already in W (par. 1), Newton clearly disagrees with Lange on the definition of Nisan 1 – it is <u>always</u> the day that begins on the night of the first visibility of the moon.

---

[24] Mishna *Rosh Hashana* (4:4) speaks about one case of confusion during the Second Temple period when witnesses were accepted in the afternoon and the priests erred in signing at the Mincha service. As a consequence, the afternoon testimony was ruled out afterwards, but Lange could have assumed that the above episode, as well as many other outstanding cases recorded in the Talmud, must be related to Jesus' life.
[25] See Lange, pp. 403-404. He apparently took this notion from John Selden, *De anno civili*, pp. 4-10.
[26] Note that the author of Yahuda 24E found the moon to be 16°54' from the sun at sunset on March 20, 33 AC, which would be satisfactory for his (Lange's) purpose.
[27] This retroactive assignment of Rosh Chodesh, generally never permitted by Rabbanim, was mentioned by Maimonides in *Sanctification of the New Moon* [III:15], saying that the court must accept witnesses retroactively, even a week or two after the beginning of the month (sic!). One can imagine that he succumbed to the pressure of the Karaite majority in 12th century Egypt and acknowledged a certain ancient tradition.

A peculiar situation is that Lange quoted only one paragraph from Maimonides – paragraph 3 from chapter 17:

> Look at the first longitude [difference in longitudes between the sun and moon] (and the first latitude). If it comes 9° exactly or less then it is certain that the moon cannot be seen on that night in all Eretz Israel. And if the first longitude is greater than 15° then it is certain that it is possible to see it in all Eretz Israel. And if it is in between 9° and 15° you have to go and check in the visibility tables [to find] whether it is possible or not.

while the explanation – that this criterion is relevant only for the spring (December-June) semicircle – is contained in paragraph 4 from the same (17[th]) chapter (Fig.2):

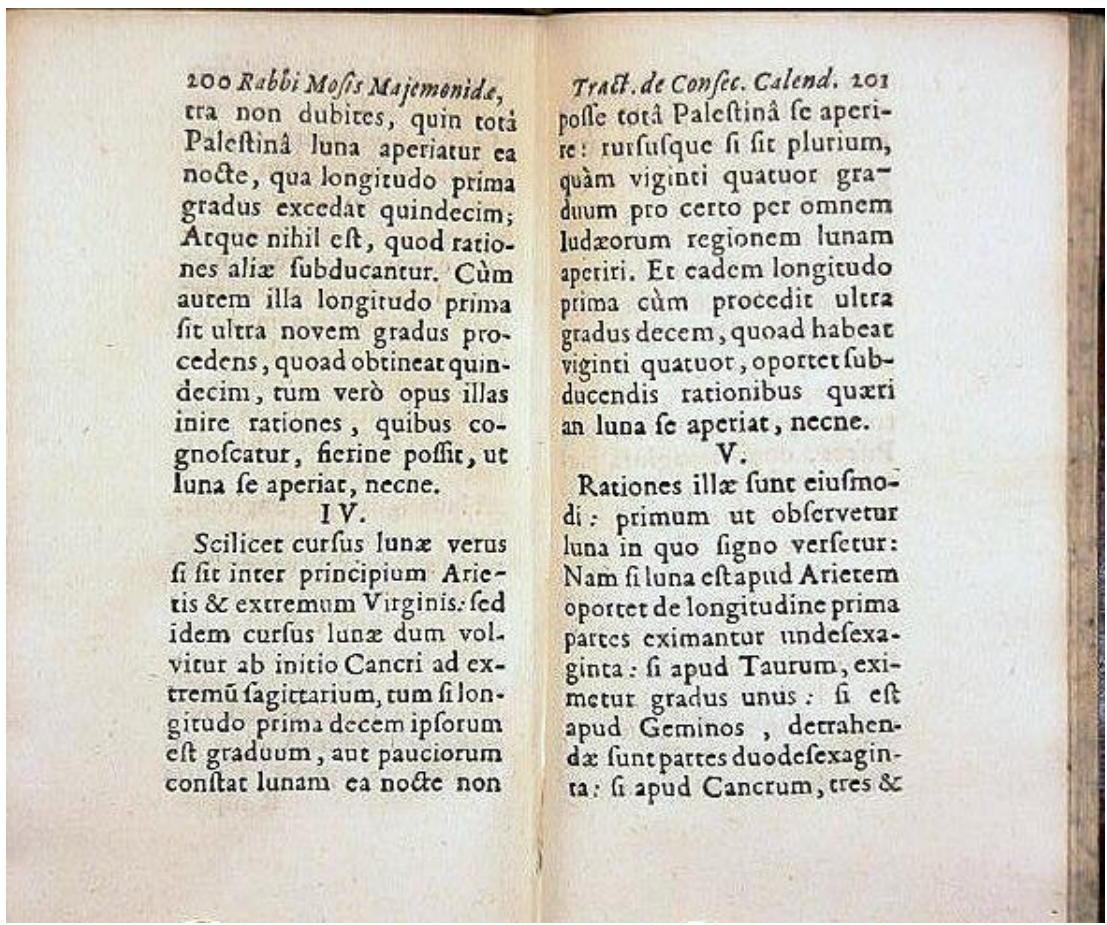

**Fig. 2.   Maimonides, Moses. *Tractatus De consecratione calendarum, et De ration intercalandi.* Paris: Petrus Promé, 1669. Translation from Hebrew into Latin by Louis de Compiègne de Veil. (Courtesy of Roger Friedman, Rare Book Studio)**

> We spoke about the place of the moon from the beginning of <u>Capricorn</u> until the end of <u>Gemini</u>. But if the moon found from the beginning of Cancer to the end of Sagittarius and the first longitude comes to 10º exactly or less then it is certain that the moon cannot be seen on that night in all Eretz Israel. And if the first longitude is greater than 24º then it is certain that it is possible to see the moon in all Eretz Israel. And if it is in between 10º and 24º you have to go and check in the visibility tables [to find] whether it is possible or not.

Translating paragraph 4, instead of the correct Capricorn and Gemini (underlined above), de Veil in the 1669 Paris edition mistakenly wrote *Aries* and *Virgo*. The error was corrected only in the next London edition, in 1683. Draft W is unaware of Maimonides' paragraph 4 and therefore does not distinguish criteria of the first visibility for the spring semicircle from the autumnal one. Hence W knew of Maimonides only from Lange.

The words in parentheses – "and the first latitude" – seemingly moved Newton to his first astronomical exegesis: in Rule 2 of W, he suggested that the actual difference in longitudes depends on the moon's latitude, and therefore moved the boundaries given by Maimonides by 2º,[28] arriving at 11º-15º for moon's southern latitude and 9º-13º for the northern latitude. With that, he confirms Lange's basic conclusion about the year of the Passion and announces April 3 (Friday), 33 AD as the only possible solution.

Draft N is already aware of paragraph 4 and avoids de Veil's error. Moreover, it contains a more powerful astronomical exegesis.

The actual ability to see the moon depends on the size (width) of the crescent. The latter is proportional to the distance between the moon and the sun. This distance is the hypotenuse in a triangle formed by the difference in longitude between the sun and moon and lunar latitude. The latter two are easily computable parameters if one knows three constants: mean motions of the sun and its anomaly, the moon and its anomaly, and the line of nodes. Newton never doubted, following Maimonides, that first century Jewish sages (priests) would know these values, and, in fact, used them.

It is clear that if people on the street were asked, they could not possibly tell the moon's distance from the setting sun and moon's latitude. In chapter 2 of the first part of his

---

[28] $2º = 5º \cdot \tan 21º$. It is unclear where the last angle might come from. Newton also quotes fractional degrees taken by Lange from Elia, who simply applied the formula for the moon's parallax, as Maimonides explained in paragraph 5 of the 17th chapter of his work.

*Sanctification of the New Moon*,[29] Maimonides asserts that witnesses in Judea were questioned about the moon's width and the direction of its horns, on which side of the sun it was at sunset (north or south), and the height of the moon above the horizon (was it at the level of the second or third floor of the house?). In chapter 6, he further states that the Sanhedrin was able to compute positions of the sun and moon and to reject outright false witnesses or be lenient toward likely true ones. Maimonides does not immediately explain to us the criteria, but in the second part of the *Sanctification,* written eight years later, he supplies a detailed discussion.[30]

Surprisingly, the criterion for visibility that Maimonides gives in chapter 17, together with the difference in longitudes (E), includes quite a different parameter, a so-called *arc of vision* (V). This is the arc of the celestial equator, which sets simultaneously with the moon. To acquire a status of "observable," it should be translated as "length of visibility period." Maimonides claimed that visibility is uncertain and needs further investigation only in the case of $9° \leq E \leq 15°$ (for the spring semi-year) and in the case of $10° \leq E \leq 24°$ (for the autumnal semi-year). In these cases, the arc of vision is checked and in the case of $9° \leq V \leq 14°$, the final criterion for the moon's visibility is $E + V \geq 22°$.

However, these criteria seem incompatible with the questions asked by the Sanhedrin, and Newton discarded the "arc of vision" as a reliable parameter and launched his own investigation.

## Newton's Criterion for the First Lunar Visibility

It is doubtful whether the *Arc of Vision* is truly "observable" and can be honestly reported. Will witnesses wait for two more hours instead of rushing to report their luck to the Sanhedrin? Accordingly, Newton never mentions this parameter and instead, in N, suggests his own, the original observable altitude **A** at sunset (**HS** – see Fig. 3). Further he takes pains to translate Maimonides' criteria, mostly expressed in ecliptic coordinates longitudes and latitudes, into the inequalities for the altitude. Deriving inequalities, he checks their consistency for the four cardinal points – the solstices and equinoxes. Of course, for the matter under discussion (time of the Passover), the vernal equinox alone represents the primary interest. In the very beginning of N, Newton formulates his "79/80-1/6" rule. The drawing in Figure 3 illuminates the matter.[31]

---

[29] Maimonides**,** Moses. *Sanctification of the New Moon*. Code of Maimonides, book 3, treatise 8. Translation by S. Gandz. Introduction by Oberman. Commentary by O. Neugebauer. Yale Judaica Series, v. 11. New Haven: Yale University Press 1952.   The first part includes chapters 1-10.
[30] In 1949 Otto Neugebauer proved that Maimonides' astronomical parameters (though not the final criteria), explained in the second part of *Sanctification*, come directly from al-Battani. See "The Astronomy of Maimonides and its Sources," *HUCA* 22, 1949.

Newton computes the moon's altitude as **A = (1 - 1/80) elongation + 1/6 latitude**, where the moon's southern latitude (as in Fig. 3) will result in a negative sign. The coefficients 1/80 and 1/6 should be *cosine* and *sine* of the angle η between the ecliptic and the vertical line at the moment of sunset. This angle is variable because the angle between ecliptic and celestial equator ε is variable: it is 0° at the solstices, +23° at the vernal equinox, and -23° at the autumnal equinox. To obtain fractions 79/80 and 1/6 for *cosine* and *sine*, angle η must fall in the range 9.3° - 9.5°. This interval is by 1.5° greater than the difference between Jerusalem's latitude (φ = 31;41°) and the inclination of the celestial pole (ε = 23;50°) at the vernal equinox. It is almost certain that Newton made the following correction, taking into account the loose connection between solar and lunar years.

Because for 13 weeks, from a solstice to an equinox, the angle ε between ecliptic and equator changes from 0° to 23;50°, Newton could assume 1.83° = 1;50° change per week. Further: all the Nisan New Moons fall within 0-14 days before the vernal equinox – on average one week. So Newton could assume 23;50° - 1;50° = 22° for angle ε between the ecliptic and equator. Accordingly: η = Jerusalem's latitude - angle ε = 31;41° - 22° = 9;41° whose *sine* is 1/6.

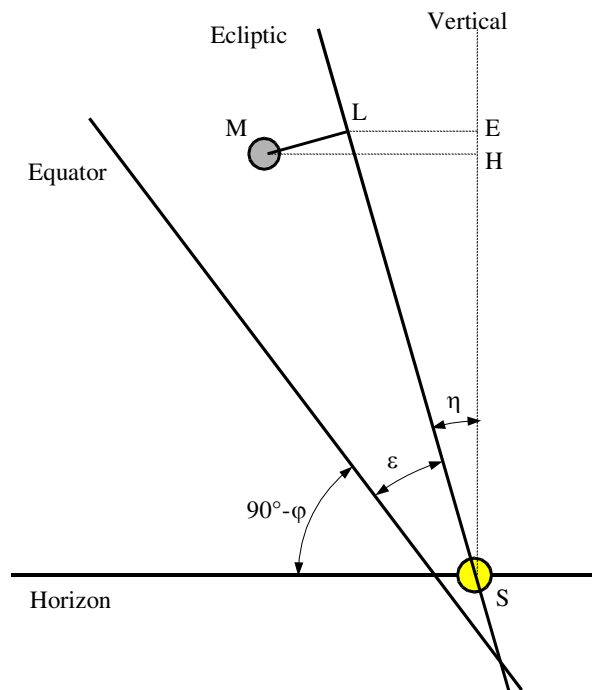

**Figure 3. Positions of the sun and the moon at sunset near the vernal equinox**

This is the picture that Newton kept in mind while looking for a general criterion in terms of altitude. He derived it from Maimonides' criteria at cardinal points. For example, he suggested that 9½° and 11° for the spring equinox (sun in Aries) came from subtracting or, respectively, adding 5/6° coming from latitude term in his formula. Therefore 79/80 of longitude and altitude must be equal to their average, 10¼°. And since Maimonides says that between Capricorn and Gemini a difference in longitude of less than 9º precludes any visibility, then altitude 10¼° can be a boundary for visibility at the vernal equinox.

---

[31]Analogous picture in Neugebauer (*ibid*, p. 354) depicts the sunset at the autumnal equinox; therefore, the ecliptic is drawn on the other side of the equator.

He supports the last value with the following example. With difference in longitude 9° and moon's latitude 5°, its elongation from the sun can be found as the hypotenuse of the right triangle:  $9^2 + 5^2 = 10;18^2$. In Gemini and Capricorn, such a moon at sunset is located perpendicularly over the sun, therefore its altitude will be equal to elongation, and therefore again close to 10¼°.[32]

In the solstices (sun in Gemini and Capricorn), the angle $\varepsilon = 0$ and $\phi - \varepsilon = 31.66°$ and therefore *cosine* $31.66° = 0.85$ while *sine* $31.66° = 0.52$. Then Newton's altitude 11°6' (=11.1°) nicely explains 10° given by Maimonides as the minimum longitude for visibility between summer and winter solstices: $11.1° = 10° \cdot 0.85 + 0.52 \cdot 5°$. Newton says here: "this is the justification for our limit of 11° or even 10½°."

At the autumnal equinox, the angle $\varepsilon = -23° 50'$ and $\phi - \varepsilon = 55°30'$ and its *cosine* = 0.57 while *sine* = 0.82. Then an altitude of 10° is close to explaining Maimonides' 24° as maximum longitude for non-visibility in case of maximal southern latitude:  $24° \cdot 0.57 - 5° \cdot 0.82 = 9.5°$.

Though Newton tried hard to reduce the problem to a single parameter – altitude – his final criteria in N still employs another parameter – elongation:

> Altitude > 10½° is required to see the Moon at night when *the distance of the Sun to the Moon* is not greater than 12°.

It is a bit disappointing that such precision was barely needed in analyzing the actual data in the tables for years 30-37, except for year 32. There he rejected April 13 as the day of first visibility, since on that day the altitude was 10° 3' while the distance between the sun and moon was 10°16'.  Besides, Newton realized that *possible* visibility does not necessarily imply *actual;* therefore, he doubled almost all of the dates in the tables.[33] It is amazing that even with that concession to nature, only year 33 in W and only year 34 in N yielded the desired result! What convinced Newton of the truth of his procedure was the unique result! True, in the second case year 34 yielded two options separated by a month. March 26, and, in case of intercalation of the 13th month, April 23, appeared to be equally plausible for the day of the Passion. This was his next puzzle to ponder.[34]

---

[32] A nuance is that in such a triangle, the *sine* of angle η between ecliptic and vertical is equal to 5 /10;18 = ½ and therefore angle η must be c. 30°, the latitude of Jerusalem. This means that the angle between the ecliptic and equator ε is 0°, which could happen only at the solstices, or, in Newton's words, in Gemini and Capricorn.

## Which Gospel Did Newton Preach?

In both drafts, Newton placed the Crucifixion on Friday, thereby disagreeing with Lange on the most important issue. Rule 4 in W leaves no doubt of his intentions: "…on the same day according to the Jews, a Friday according to the Romans, the Passion took place."[35] The same statement is repeated in rule 5 of N:

> On the first day of unleavened bread, also the eve of Nisan 15, the Lord ate the Passover with his disciples, the Roman's Thursday, and, as the same Jewish day advanced, he was crucified on Friday.

**Figure 4: Table of Paschal Crescent Visibility (Fair Copy) in Wickins' hand, with corrections by a different hand. (Courtesy of JNUL)**

[33] The doubling was unnecessary. See, for instance, "Dating the Crucifixion" by C.J. Humphreys and W.G. Waddington (note 18 above) which agrees with Newton's first (earlier) choices for years 30-36 (except for year 32, where Newton later wiped out April 13). The reason for the coincidence with the earlier choices lies in the improved criteria for visibility that allowed for sighting a thinER crescent in the olden days.

[34] Modern researchers consider a wider historical interval and also include years 27 (April 11) and 30 (April 7) as possible options. See Humphreys and Waddington (note 18 above) and also two papers by Bradley E. Schaefer, "Dating the Crucifixion," Sky and Telescope, April 1989, p. 374, and "Lunar Visibility and the Crucifixion" in Q.J.R.Astr. Soc. 31 (1990), pp. 53-68, or the article by Eduardo Vila Echagüe "¿Cuándo murió Jesucristo?" (June 2000) at <http://www.geocities.com/edovila/Crist/MuerJesu.html>.

[35] In the passage "which is when the Lord was allowed to suffer" (end of W) 'which' refers to the year 33, not to the day (Thursday).

A remarkable feature of W is the superscript that Newton inserted above every date (see Fig.4). (These superscripts became the principal dates in N.) They emphasize the fact that in W, Newton first added 13 days to the eve of the first visibility ('Vesper Primi Nisan'[36]); therefore, he believed with Lange that the Last Supper and arrest happened on the eve of the official Nisan 14. But in W he refers to a real Paschal meal partaken by Jesus; therefore, he accepted Lange's idea about "double standards" in the behavior of Sadducees, who could celebrate their Passover a day earlier.

The superscripts inserted above all the dates in W in Newton's hand signify a change at a later time to the idea that Thursday was the eve of Nisan 15 and the Passion occurred on Friday in the wake of the festival, though before Shabbat, in parallel with the opinion of the synoptic gospels.[37] This change was reflected in N. Since St. John occupied a special place in Newton's theology, this remarkable switch of allegiance could not possibly have been easy for a person of Newton's temperament. Arguably, this could be at the bottom of several puzzling facts of Newton's life: his inexplicable break with Wickins in 1678-83 [38] and undoubtedly the late appearance of the *Observations,* where the opening phrase in chapter 11 is: "it is assumed *without saying* (sic!) that the Passion occurred on Nisan 14."[39] The reason for such a change of mind could have been both theoretical and practical. Finding an extra (fifth) Passover in the gospel narrative, and keeping year 30 as a starting point, Newton decided, for the Passion, in favor of year 34.[40] This was strengthened by pure empirical argument, quoted in the beginning of the paper (the barley could be ripe on April 13 but not on March 27!) which, with all probability, came from his early days on the farm at Woolsthorpe, where he surely could have acquired some knowledge of the proper time for grain cultivation.[41] After reading Maimonides, he could discard Lange's idea of an individual Passover. Therefore, Thursday was discarded in favor of Friday, and Nisan 14 in favor of Nisan 15. Both were allowed by the late Passover in year 34.

[36] When the word comes before a date expressed in genitive case ('vesper Primi Nisan', 'vesper Paschalis') it means 'eve', which is the evening of the previous day. But if the word comes after a date, usually in ablative case ('seu 14 die Nisan Vesperi', 'et inde 14 Nisan vespere') it has the conventional meaning of 'evening' (of the same day). The Jewish day is from sunset to sunset.

[37] Mark [15:42] and Luke [23:54] explicitly point to Friday—the day before Shabbat (Matthew seemingly did not specify it). Taking the Last Supper to be a Paschal meal on the night before the Passion, this implies that Friday was also Nisan 15. See Humphreys & Waddington on variant interpretations of the Synoptics.

[38] See Westfall, 342.

[39] I. Newton. *Observations upon the Prophecies of Daniel and the Apocalypse of St. John* (1733). See also the English transcription <www.newtonproject.sussex.ac.uk>.

[40] The idea that five Passovers can be found in the gospels was first introduced by Gerhard Mercator (1569). It was upheld by Scaliger (1583, 1598) and Lange, pp. 429-433. Not all scholars supported this idea: e.g., Petavius (1627, pp. 445-451) counted only three Passovers in the gospels. We owe this information to Philipp Nothaft.

[41] See Westfall, 63, for peculiar details. Even taking it *cum grano salis* one cannot see how Pratt's statement – that Newton's argument of "ripeness of corn" did "not pass historical judgment" – can be taken seriously.

**Figure 5. Table of Paschal Crescent Visibility (Rough Draft). (Courtesy of JNUL)**

This approach precipitated another trap. Year 34 does not forbid Passover eve from falling on March 25, in case of poor visibility on March 10 and the subsequent announcement of March 11 as the first day of the month. The resulting lack of uniqueness of the solution might be a bit irritating for a physicist looking for a unique solution! A question mark [?] introduced in the next to the last column of the table in N on the line for 34 AC (see Fig. 5) suggests a new stage in Newton's mental process. The philosopher, arguing against Cartesian *necessity* and defending God acting *out of freedom* in Nature, could not bear such a freedom in history.

### *Ex Ungue Leonem*

The posthumously published *Observations* gave Newton's final solution of the problem. To preserve St. John's authority, while retaining a unique solution within year 34, Newton summons a feature of the Jewish calendar that could not possibly have been in use in the first century AD: postponements, in Hebrew – *dekhiyot*. The essence of dekhiyah YaĤ (literally "18") is to postpone Rosh Hashana to the next day if Molad[42] Tishrei falls after noon (18h according to the Jewish reckoning). The essence of dekhiyah ADU-Rosh is to postpone Rosh Hashana to the next day if Molad Tishrei falls on Sunday, Wednesday, or Friday. (This, in turn, immediately restricts the days on which Passover can fall to Sunday, Tuesday, Thursday, and Saturday.) Formally, Newton's logic is faultless – the Jewish calendar, in the words of its contemporary expositors, *was given at Mount Sinai*, i.e., it is eternal and unchangeable.[43]

---

[42] Jewish calendar conjunction, annually computed from a so-called "Molad BAHARAD" of 3761 BC by adding each year an appropriate number of mean months of 29 d 12 h 44 min 3.3 s. See Stern, *Calendar and Community*, 191-2, or A. Belenkiy "A unique feature of the Jewish calendar - Dekhiyot" (note 19).

Newton does not quote any more from Lange, Maimonides, or Elia. Neither does he care about their explanations on what each type of postponement means. He totally disregards a somewhat confused Talmudic explanation about the 18 hours.[44] He knows that if the Jews had followed any logic, "18 hours" could mean only the time interval following a conjunction:

> …and because the first appearance might usually be about 18 hours after the true conjunction, they therefore began their month from the sixth hour [**161**] at evening, that is, at sun set, next after the eighteenth hour from the conjunction. And this rule they called י"ח designating by the letters י and ח the number 18.

Newton already had chosen year 34 and not year 33. However, the latter year has Nisan 14 on Friday, and it is impossible that in two consecutive years the holiday will fall on the same weekday. As a rule, a difference of 4-5 days is expected, because an average lunar year consists of 354.3 days. Then Newton makes a true *salto mortale* – he finds the missing (to complete seven days) 2.7 days: 1.5 days by assuming an additional (thirteenth) month in the Jewish year 33-34, and one more day due to a historically questionable postponement "BDU-Passover," which prevents Passover from falling on Monday, Wednesday and Friday by postponing it to the next day (via adding an extra, 30th day in Adar). In addition:

> [**162**]…Computing therefore the new moons of the first month according to the course of the moon and the rule י"ח and thence counting 14 days, I find that the 14th day of this month in the year of *Christ* 31, fell on tuesday *March* 27; in the year 32, on sunday *Apr*. 13; in the year 33, on friday *Apr*. 3; in the year 34, on wednesday *March* 24, or rather, for avoiding the Equinox which fell on the same day, and for having a fitter time for harvest, on thursday *Apr*. 22, also in the year 35, on tuesday *Apr*. 12, and in the year 36, on saturday *March* 31.

---

[43] The discussion of *dekhyiot* in connection with the date of the Passion goes back to Paul of Burgos (Samuel Ha-Levi, 1351-1435) and his *Additiones* (1429) to Matthew 26. The idea was picked up by Paul of Middelburg (*Paulina*, 1513), later by Sebastian Münster (*Kalendarium Hebraicum*, 1527), and finally by Scaliger (*De emendatione temporum*, 1583). They were criticized by Petavius (1627, II, pp. 430-440), who realized that the postponements were introduced only much later. (This fact we owe to Philipp Nothaft). Newton owned the books of Münster, Scaliger and Petavius.

[44] A Talmudic statement: "The father of Rabbi Simlai asked Shmuel: 'Do you know the difference between whether the Molad is before or after noon?' … Rabbi Zeira said: 'This is what the father of Rabbi Simlai meant: if the Molad comes after noon, we do not see it the same evening; but if before, we do see it'…" (Talmud Bavli, Rosh Hashana 20b) was usually taken by exegetes as THE reason for the Dekhiyah YaĤ because the Jewish day begins exactly 18 hours earlier than noon. See Stern, *Calendar and Community*, 192, or A. Belenkiy, "A unique feature of the Jewish calendar − Dekhiyot" (note 19).

But because the 15th and 21st days of *Nisan*, and a day or two of *Pentecost*, and the 10th, 15th, and 22d of *Tisri*, were always sabbatical days or days of rest, and it was inconvenient on [**163**] two sabbaths together to be prohibited burying their dead and making ready fresh meat, for in that hot region their meat would be apt in two days to corrupt: to avoid these and such like inconveniences, the *Jews* postponed their months a day, as often as the first day of the month *Tisri*, or, which is all one, the third of the month *Nisan*, was sunday, wednesday or friday: and this rule they called אד"ו by the letters א, ד, ו signifying the numbers 1, 4, 6; that is, the 1st, 4th, and 6th days of the week; which days we call sunday, wednesday and friday. Postponing therefore by this rule the months found above; the 14th day of the month *Nisan* will fall in the year of *Christ* 31, on wednesday *March* 28; in the year 32, on monday *Apr.* 14; in the year 33, on friday *Apr.* 3; in the year 34, on friday *Apr.* 23; in the year 35, on wednesday *Apr.* 13; and in the year 36, on saturday *March* 31.

With the understanding that 'saturday' could be also a viable option, Newton-the-historian makes a careful historical analysis of the political situation in Judea, finally discarding years 35 and 36. The whole argument, however, is highly doubtful: the first references to postponements appeared only in rabbinical literature of the late third century AD. Newton certainly knew the weakness of his own general argument where "two sabbaths together to be prohibited burying their dead and making ready fresh meat" – the modern set of postponements do not prevent Passover from falling on Sunday, although it is prevented from falling on Friday and that was what he actually wanted!

### The Tables

The tables (Figs. 3 & 4) compute several solar and lunar parameters for years 30-37 for the night of the first visibility in Jerusalem of the last-winter or first-spring moon (Table 2).

| AC March | | Long. Sun | Long. Moon | Dist. M from S. | Lat. Moon | Alt. Moon at | Eve of Nisan 1 | Passover eve or Nisan 15 | |
|---|---|---|---|---|---|---|---|---|---|
| 30 | 23 | 0. 0. 45 | 0.10.33 | 9. 48 | 4.0 S | 9.48 | Mar 24, 25 | Apr 7,8 | Fr,Sa |
| 31 | 12 | 11.19.51 | 11.28.15 | 8. 24 | 3.35 S | 7. 40 | Mar 13,14 | Mar 27,28 Tu,We | |
| 32 | 30 | 0. 8. 6 | 0.18.22 | 10. 16 | 0.33 S | 10. 3 | Mar 31 | Apr 14 | Mo |
| 33 | 20 | 11.28.10 | 0.15.4 | 16. 54 | 0.49 N | 16. 50 | Mar 20, 21 | Apr 3,4 | Fr,Sa |
| 34 | 9 | 11.17.10 | 11.24.15 | 7. 5 | 0.44 N | 7. 6 | Mar 10,11 | Mar 24,25 We,Th | |
| 35 | 28 | 0.5.28 | 0.12.28 | 7. 0 | 3.39 N | 7. 30 | Mar 29 | Apr 12,13 Tu,We | |
| 36 | 17 | 11.25.30 | 0.7.40 | 12. 10 | 4.21 N | 12. 34 | Mar 17, 18 | Mar 31,32 Sa,Su | |
| 34 Apr 8 18h 16m | | 0.16.20 | 1.3.10 | 16. 50 | 3.44 N | 17. 15 | Apr 8, 9 | Apr 22,23 Th,Fr | |

**Table 2. Yahuda 24E's table displayed in Fig. 5.**

They include the sun's and the moon's true longitudes and the moon's latitude, computed for 3 p.m. London time, roughly equivalent to 6 p.m. Jerusalem time, a sunset time in the spring.[45]   In both drafts, there is also a column labeled "Distance between the moon and the sun" that contains the difference between their longitudes. In N there is an additional column with the moon's altitudes at sunset. These tables, in principle, can provide a clue for dating W and N, if their source were identified.

### a. General state of pre-Newtonian astronomy

Throughout the 17[th] century, the theory underlying the motion of the celestial bodies was rapidly evolving. Since the 1660s, Kepler's *Rudolphine Tables* (1627) were considered the best for computing the positions of the planets, but not for the moon, where many different theories vied for the laurel wreath. In his *Almagestum Novum,* published in 1651, Giovanni Battista Riccioli (1598-1671), in addition to the theory of Johannes Kepler (1571-1630), describes lunar theories by Philip van Lansbergen (1561-1632), Godfroy Vendelin (1580-1667), Johannes Fabricius (1587-1616), Albert Curtz (1600-1671), and Ismael Boulliau (1605-1644), who published extensively during the mid-17[th] century on the Continent, while at the same time, in England Vincent Wing (1619-1668) and Thomas Streete (1622-1689) suggested rival geometrical theories, and Jeremiah Horrox (c. 1617-1641) developed his own theory, published by John Wallis only in 1673 with John Flamsteed's 1672 tables,   which were based on Horroxian theory. These tables contained some errors; in 1681, Flamsteed, then Astronomer Royal, published corrected tables. Newton's own lunar theory, based on the Universal Law of Gravity conceived in *Principia*, was developed much later and was first published in 1702 by David Gregory.[46]

### b. Comparison of Yahuda Table with Those of Major Astronomers

*Almagest*,[47] al-Battani,[48] the *Alphonsines*, the *Prutenics*, Tycho's *Astronomiae Insturatae Progymnasmata,*[49] Kepler's *Rudolphines*,[50] Horrox's *Opera Posthuma,*[51] and Streete's

---

[45] The single exception is the year 34 AC entry, where the time is specified as 18h 16m, likely a direct reference to Jerusalem local time of a sunset in April in Jerusalem.
[46] See extensive review in C. Wilson. "Predictive Astronomy in the Century after Kepler" in: General History of Astronomy. Cambridge, 1989, vol. 2A, pp. 161-206.
[47] Lars Gislén. *Astromodels* Freeware. http://www.thep.lu.se/~larsg/download.html. Based on Ptolemy's *Almagest*, *The Alphonsine Tables*, Erasmus Reinhold's *Tabulae Prutenicae* of 1551, Thomas Streete's *Astronomia Carolina* of 1665, and Tycho Brahe's models in *Astronomiae Instauratae Progymnasmata.*
[48] Al-Battani, Muhammad ibn Jabir. *Opus Astronomicum.* Translation from Arabic to Latin and commentary by C.A. Nallino. Part 2. (Milano 1907); reprint (Georg Olms Verlag, Hildesheim, N.Y. 1977), p. 72.
[49] True, Harrison does not list Kepler and Tycho in Newton's library, but Barrow's library did contain them; see J.R. Harrison. *The Library of Isaac Newton.* Cambridge Univ. Press 1978 and note 65 below.
[50] Kepler, Johannes. *Joannis Kepleri astronomi, Opera omnia Volumen VI / ed. Ch. Frisch.* De Tabulis Rudolphinis. Gallica. http://gallica.bnf.fr/

*Astronomia Carolina* are possible sources for Yahuda 24E. Instead of comparing them vs. Yahuda 24E's tables directly, we compared all with modern theory[52] (see table 3, where the differences are expressed in minutes of arc).

| Date | Yahuda | Alma-gest | Al-Battani | Alphon-sines | Prutenics | Tycho | Rudol-phines | Streete | Flam-steed | Newton 1702 |
|---|---|---|---|---|---|---|---|---|---|---|
| Mar 23, 30 | 6 | -20 | -90 | 38 | -22 | -6 | 4 | 7 | -1 | -3 |
| Mar 12, 31 | 11 | -17 | -89 | 38 | -19 | -8 | 5 | 5 | 0 | -3 |
| Mar 30, 32 | 8 | -24 | -89 | 38 | -25 | -4 | 4 | 8 | -1 | -3 |
| Mar 19, 33 | 10 | -20 | -89 | 38 | -22 | -6 | 4 | 6 | -1 | -3 |
| Mar 20, 33 | 10 | -21 | -89 | 38 | -22 | -6 | 4 | 6 | -1 | -3 |
| Mar 9, 34 | 10 | -18 | -89 | 38 | -20 | -9 | 4 | 4 | 0 | -3 |
| Mar 28, 35 | 9 | -24 | -90 | 38 | -25 | -5 | 3 | 7 | -1 | -3 |
| Mar 17, 36 | 9 | -21 | -90 | 37 | -23 | -7 | 4 | 5 | -1 | -3 |
| Apr 4, 37 | 10 | -27 | -89 | 38 | -28 | -4 | 3 | 9 | -1 | -3 |
| *Apr 8, 34* | *8* | *-27* | *-89* | *38* | *-28* | *-3* | *3* | *10* | *-1* | *-3* |
| | | | | | | | | | | |
| Std. Dev. | 1.4 | 3.4 | 0.2 | 0.2 | 3.2 | 1.7 | 0.5 | 1.8 | 0.3 | 0.3 |
| ρ (X,Yahuda) | | 0.20 | 0.38 | -0.23 | 0.24 | -0.40 | 0.23 | -0.39 | 0.18 | -0.01 |
| σ_ ρ (X, Yahuda) | | 0.32 | 0.29 | 0.32 | 0.31 | 0.28 | 0.32 | 0.28 | 0.32 | 0.33 |

**Table 3. Comparison of solar positions of Yahuda 24E and others vs. modern theory**

Apart from the absolute size of errors coming from the different values assigned to the sun's mean motion, Yahuda 24E's solar positions show sample standard deviation = 1.4 of the same order as those of Tycho and Streete, and higher than those of 1681 Flamsteed and 1702 Newton. The correlation coefficients in the sun's longitude between Yahuda 24E and all other astronomers' tables are very low. Even the two largest in absolute value correlation coefficients – Tycho (ρ = -0.4) and Streete (ρ = -0.39) – are significant only at the 20% level! However, it is not that easy to decide on which correlations are significant, since the sample of 10 is rather small and one or two accidental computational errors could reverse the picture. Tycho and Streete were excluded by detailed comparison of Yahuda 24E with tables prepared by Lars Gislen for Tycho and Streete. Poor correlation between Yahuda 24E and the Rudolphines seemed to exclude Kepler.

---

[51] Harrison, p. 163, entry 808. Horrocks, Jeremiah. *Opera posthuma*; viz. *Astronomia Kepleriana.*, defensa & promota… J. Flamstedii, de temporis aquatione diatriba. Numeri ad Lunae theoriam Horroccianam. In calce adjiciuntur, nondum editae…4° Londini, 1678. We used two of Nick Kollerstrom's programs: Flamsteed's 1681 Horroxian Lunar Theory. A Computer Simulation at <http://www.ucl.ac.uk/sts/nk/zip/flam1.zip> and Newton's 1702 Theory of the Moon's Motion. A Computer Simulation at <http://www.ucl.ac.uk/sts/nk>.

[52] For the sun: Pierre Bretagnon and Jean-Louis Simon, *Planetary Programs and Tables from 4000 to +2800*. Willmann-Bell Inc., 1986. For the moon: Michelle Chapront-Touzé and Jean Chapront, *Lunar Tables and Programs from 4000 B.C. to A.D. 8000*. Willmann-Bell Inc., 1991.

| Date | Yahuda | Alma-gest | Al-Battani | Alphon-sines | Prutenics | Tycho | Horrocks | Streete | Flam-steed | Newton 1702 |
|---|---|---|---|---|---|---|---|---|---|---|
| Mar 23, 30 | 10 | -40 | -81 | 22 | -52 | -3 | 11 | 69 | 10 | -8 |
| Mar 12, 31 | 10 | -41 | -82 | 10 | -46 | 2 | 19 | 34 | 10 | -14 |
| Mar 30, 32 | 19 | -37 | -78 | 11 | -39 | 2 | 17 | 22 | 10 | -21 |
| Mar 19, 33 | 34 | -18 | -61 | 21 | -15 | 8 | 12 | 18 | 11 | -25 |
| Mar 20, 33 | 37 | -42 | -84 | -2 | -28 | 7 | 15 | 18 | 14 | -22 |
| Mar 9, 34 | 31 | -34 | -77 | 6 | -18 | 7 | 5 | 20 | 6 | -22 |
| Mar 28, 35 | 58 | -32 | -74 | 14 | -17 | 7 | 10 | 23 | 10 | -18 |
| Mar 17, 36 | 26 | -36 | -76 | 21 | -33 | 10 | 3 | 24 | 12 | -14 |
| Apr 4, 37 | 32 | -25 | -65 | 37 | -27 | 0 | 15 | 16 | 16 | -11 |
| *Apr 8, 34* | *48* | *-41* | *-83* | *-3* | *-20* | *7* | *17* | *18* | *15* | *-20* |
| | | | | | | | | | | |
| Std. Dev. | 15.4 | 7.7 | 7.8 | 12.1 | 12.8 | 4.2 | 5.4 | 16.0 | 2.9 | 5.5 |
| $\rho$ (X,Yahuda) | | 0.25 | 0.17 | -0.27 | 0.85 | 0.59 | -0.10 | -0.61 | 0.26 | -0.47 |
| $\sigma\_\rho$ (X,Yahuda) | | 0.31 | 0.32 | 0.31 | 0.09 | 0.22 | 0.33 | 0.21 | 0.31 | 0.26 |

**Table 4. Comparison of the lunar positions of Yahuda 24E and others vs. modern values**

Not only do Yahuda 24E's lunar positions have large differences with the modern values, stretching within the range between +10' and +58', but the sample variance is too great (see Table 4). Even the *Almagest* performs better in this respect. The correlation coefficients in the moon's longitude between Yahuda 24E and all others are very low. The notable exception is the *Prutenics*, which shows a very high correlation ($\rho = 0.85$) significant at < 0.0001 level! The next two, by comparison, are Tycho ($\rho = 0.7$) and Streete ($\rho = 0.7$), significant only at the 2% level. However, the *Prutenics* disagree with Yahuda 24E in absolute values too much to be seriously considered as its source.

### b. Newton's contemporaries

Special attention must be given to Newton's older contemporaries: Nicholas Mercator (born Kauffmann, 1620-87) and Vincent Wing.

### Nicholas Mercator

Westfall writes [p. 258]:

> A note dated 1673 in the Appendix to Edward Sherburne's *The Sphere of Manilius* (1675), stated that Newton had a treatise on dioptrics ready to the press, "and divers Astronomical Exercises, which are to be subjoined to Mr Nicholas Mercator's *Epitome of Astronomy* and to be printed in Cambridge." He added that Newton also planned to publish a general analytic method based on infinite series for quadrature of figures, centers of gravity, volumes, surfaces and rectifications. Whatever the source of Sherburne's information, we know nothing more about the astronomical exercise. They did not appear in Mercator's book when he published it in 1676, though a reference to Newton, who had shown the author a very elegant hypothesis on moon's libration, establishes that the two had met.

Nicholas Mercator, once a lecturer at the University of Copenhagen (1648-54), resided in London in 1658-82. His "Epitome" (*Institutionum Astronomicarum Libri Duo,* 1676) contains handy solar tables based on Tycho and Kepler and lunar tables based on Tycho.[53] Several arguments can be adduced against this book as a source for Yahuda 24 E.

First, Mercator arranged the tables in a strange fashion; he used the Julian (Scaliger's) period as an epoch (year -4712) and "decimals of a circle" instead of traditional signs, degrees, minutes and seconds, while Yahuda 24E uses traditional notation and reference point. The matter of conversion forth-and-back might be an unpleasant one.[54] Second, if Newton had used them, Yahuda 24E's table would be close to Tycho or Kepler, which is not the case, as we saw earlier. This means Mercator is an unlikely source for W.

On the other hand, Newton hardly would discard Mercator unless his own work had already been completed some time earlier. This suggests the table in the W was composed before 1676. To find the precise time of its composition, we must discover the original source.

### Vincent Wing

Harrison lists Wing's 1651 *Harmonicon Coeleste* as well read and the 1669 *Astronomia Britannica* as heavily perused and dog-eared, in Newton's manner,[55] and it is most probable that Newton used either for his solar data (see Table 5). In eight cases out of ten, Wing's 1669 procedure gives exactly the same result for sun's longitude as Yahuda 24E, if rounded to a whole number of arcmins. In one case, AC37/4/04, the difference is 1', while in the other case, AC30/3/23, it is 4' – but in the latter case the 45' in Yahuda 24E could be just a slip of the pen for 49', in which case the match with Wing 1669 is perfect.[56] Though the 1651 positions deviate from Yahuda 24E more seriously – three times by 1' and twice by 2' – they are still within an error-margin, and the 1651 book can't be discarded outright.

---

[53] This solves the puzzle we proposed in "History of One Defeat" (note 3) – how Tycho's value for the tropical year was picked up by Newton in 1700. He took it from Mercator's book!
[54] We owe this remark to Lars Gislen.
[55] Harrison, *op. cit.*, p. 300, entries 1742 and 1743.
[56] Replacing 45' with 49' gives also an almost perfect correlation between the Yahuda 24 E and *Rudolphines* solar positions, which is not surprising, since Wing's 1669 procedure approximates Kepler's area law very closely. See "Predictive Astronomy in the Century after Kepler" (p. 178).

| Table Date yy-mm-dd | London Time h:mm | Sun's Longitude | | | Moon's Longitude | | | Moon's Latitude | | |
|---|---|---|---|---|---|---|---|---|---|---|
| | | Yahu-da 24 | Wing 1669 | Wing 1651 | Yahu-da 24 | Wing 1669 | Wing 1651 | Yahu-da 24 | Wing 1669 | Wing 1651 |
| 30-03-23 | 15:00 | 0°45' | 0°49' | 0°50' | 10°33' | 10°43' | 10°42' | -4°00' | -3°59' | -3°58' |
| 31-03-12 | 15:00 | 349°51' | 349°51' | 349°51' | 358°15' | 358°29' | 358°31' | -3°35' | -3°37' | -3°36' |
| 32-03-30 | 15:00 | 8°06' | 8°06' | 8°08' | 18°22' | 18°28' | 18°33' | -0°33' | -0°30' | -0°30' |
| 33-03-19 | 15:00 | 357°11' | 357°11' | 357°11' | 0°06' | 0°04' | 0°13' | -0°26' | -0°28' | -0°28' |
| 33-03-20 | 15:00 | 358°10' | 358°10' | 358°11' | 15°04' | 14°57' | 15°07' | 0°49' | 0°50' | 0°50' |
| 34-03-09 | 15:00 | 347°10' | 347°10' | 347°11' | 354°15' | 354°15' | 354°20' | 0°44' | 0°40' | 0°40' |
| 35-03-28 | 15:00 | 5°28' | 5°28' | 5°29' | 12°28' | 12°01' | 12°02' | 3°39' | 3°35' | 3°33' |
| 36-03-17 | 15:00 | 355°30' | 355°30' | 355°30' | 7°40' | 7°49' | 7°50' | 4°21' | 4°21' | 4°20' |
| 37-04-04 | 15:00 | 12°46' | 12°45' | 12°46' | 14°56' | 14°49' | 14°52' | 5°00' | 5°00' | 5°00' |
| 34-04-08 | 15:00 | | 16°20' | 16°22' | | 33°10' | 33°02' | | 3°44' | 3°44' |
| 18:16 Jeru- | 15:11 | 16°20' | | 16°22' | 33°10' | | 33°08' | 3°44' | | 3°44' |
| salem Time | 15:14 | | 16°21' | | | 33°02' | | | 3°44' | |

**Table 5. Comparison of solar and lunar positions of Yahuda 24E and
Wing's 1669 and 1651 books**

The moon could demonstrate the true source, but the moon's differences between Yahuda 24E and Wing are much greater than for the sun, amounting to 14'-16' (in year 31) and even to 26'-27' (for year 35) and obviously need an explanation.

According to Lars Gislen[57] there are at least three options that Newton could have used for his moon calculations. First, he could have skipped the *annual equation*. Sometimes calculating according to Wing's procedure without that equation seems to give a closer fit to Yahuda 24E's data. (True, the coefficient in the annual equation was widely debatable at that time -- Tycho argued for 4.5'; Kepler argued for 11'; Boulliau neglected it altogether; and Wing in 1669 sided with Tycho.[58] Newton could have chosen another figure than Wing's.) Second, Newton could have chosen between Wing's orbital and ecliptic longitude for the moon. Generally, it seems that Wing's orbital longitudes give a better fit to Yahuda 24E, but not always. Third, Newton could have simplified the interpolation or calculation scheme. The evection in Wing 1969 is interpolated in a double-entry table and the result of the interpolation can differ by up to 30"-50" from an exact calculation. Also, the sheer number of different steps in the moon calculations can cause truncation errors.[59]

All this, however, can explain a discrepancy between Wing's 1669 procedure and Yahuda 24E of 11-13' maximum, but not of 27' unless the latter author did his computations

---

carelessly or just neglected all the terms in the lunar longitude's expansion beyond the equation of time and evection. The variation, *38' sin 2D*,[60] must be the major among the neglected terms. Since elongation D in Yahuda 24E does not exceed 17º, the *sin 2D* for each entry is not greater than 0.56, and therefore the variation does not exceed 21'. Combined with several minor terms in lunar longitude expansion, also neglected, it could produce, in principle, a discrepancy as large as 27'.

Since we did not achieve an exact match, it is worthwhile to search for extra clues in Wing's books owned by Newton.

### 1651 *Harmonicon Coeleste*

A copy of Wing's 1651 book from Newton's library is kept in the Butler Library at Columbia University, New York. There are about 20 marginal notes in Newton's hand. The notes don't show any particular interest in the moon, but there is a short table for the positions of Jupiter. The manuscript attached to the book, a "loose leaf with Newton's handwritings on both sides," in Harrison's words, was thought to be missing.[61] At our request, librarians at Columbia University's Butler Library searched for it, and on Dec.1, 2006, finally recovered the page. Unfortunately, the leaf deals solely with general chronology and is not concerned with the Passion.[62]

### 1669 *Astronomia Britannica*

According to Whiteside,

> in autograph notes made about 1670 on the rear endpapers of his copy (Trinity College, Cambridge, NQ 18.36) of Vincent Wing's *Astronomia Britannica*, London, 1669, Newton explains the disturbance of the moon's orbit from its theoretical elliptical shape through the action of the solar vortex (which 'compresses' the terrestrial one bearing the moon by about 1/43 of its width).[63]

We did not have an opportunity to examine these and other notes Newton made in this book. The conjecture is that Newton corrected Wing's parameter, say, eccentricity of the moon's ellipse, by the above mentioned "1/43". Such a correction would affect not just the minor terms in the expansion f the moon's longitude, but the major terms as well! The first, equation of center, is *(2e + ε ) sin M*,[64] where e is eccentricity of moon's ellipse, ε is the radius of the circle on which one lunar focus moves around the sun, and M is Moon's mean anomaly. Since Wing assumed ε = 0.02158 and e = 0.04315, or equivalently, e = 150', the 1/43 part of 2e alone might lead to an increase in the coefficient by up to 7'.

---

[60] See "Predictive Astronomy" (p. 195).
[61] Butler librarian Consuelo W. Dutchke, in a private communication of November 15, 2006, confirmed that "the manuscript leaf formerly was tipped in Columbia's copy of Vincent Wing's *Harmonicon Coeleste* (London 1651) and is indeed missing from the volume."
[62] Private communication from another Butler librarian, Jennifer B. Lee, on December 8, 2006.
[63] D.T. Whiteside, "Newton's Early Thought on Planetary Motion, A Fresh Look," British Journal for the History of Science, **2**, pt.2, Dec. 1964, 117-37, ftn. 11.

### 1651 or 1669 Book? – a nuisance

The lunar position for Apr 8, 34 AC, which stands alone in W, suggests a major conundrum. It is computed, not for *3 p.m.* London time, but for *18:16* – presumably Jerusalem time – the time of the supposed sunset in Jerusalem. This would correspond to different London times in two Wing's books, since Wing assigned different longitudes for Jerusalem in 1651 and 1669. It is probable that the author of W computed the solar position for AC34/4/08, not for 3 p.m. London time, but for 3:11 p.m. (1651) or 3:14 p.m. (1669). The 1651's position at 33º 08' is much closer to Yahuda's 33º 10' than 1669's 33º 02'. But again, it does not fit precisely. If, however, we set aside the mysterious 18:16 time for year 34 and come back to the standard 3 p.m. London time, then the 1669 book matches Yahuda MS 24E precisely.

### Newton's Views on the Passion: Three Stages

Yahuda 24E produces several surprising discoveries regarding Newton's views on the Passion. The posthumously published *Observations* revealed just the tip of the iceberg. Compressed into a paper, the whole account develops like a detective story that unveils three stages in Newton's theological thought:

1) his first guides for the Biblical account became Villum Lange and St. John. Newton's short-lived romance with Lange's book resulted in a thorough familiarity with the Jewish calendar traditions and some astronomy;

2) his thorough acquaintance with Maimonides led to his parting with Lange and to new insights into the first lunar visibility problem. The late maturity of the corn (barley) and hence the necessity for a late Passover two years prior to the Crucifixion, caused him to decide in favor of year 34 and the synoptic gospels' account of the Passion;

3) his study of the postponements of the Jewish holidays led to the redemption of St. John's authority.

---

[64] See "Predictive Astronomy", pp. 195-7.

Choosing year 34 for purely practical considerations, Sir Isaac later defends St. John's version vs. the synoptic gospels on the ground that the first century Jewish calendar had some peculiar features (postponements) which it actually acquired much later. Though this trick became easy prey for modern researchers[65] and his date, year 34, subsequently was considered the aberration of a weary mind and neglected by historians, Newton's chain of reasoning, set forth in Yahuda MS 24E, would still be unbroken if it were reestablished on the authority of the synoptic gospels. One can recall another passage from chapter 11 of the *Observations*:

> **[p. 151]…**This is the beginning of his miracles in *Galilee*; and thus far *John* is full and distinct in relating the actions of his first year, omitted by the other Evangelists. The rest of his history is from this time related more fully by the other Evangelists than by *John*; for what they relate he omits.

Whether Sir Isaac would himself agree on such a revision (if he were convinced that postponements are irrelevant in the Temple period) and side again with the synoptic Gospels is doubtful – his *faithfulness* to St. John was equal to his *faith* in the Scripture.

## Dating Yahuda 24E

Wickins' handwriting establishes 1683, or even 1678, as a firm upper time bound for W. On the whole this fact is not helpful, since Yahuda 24E belongs to the late 1660s or early 1670s. One of two books by Vincent Wing appears to be Newton's source for computing the lunar and solar positions in years 30-37. There remains uncertainty as to which of two Wing's books – 1651 or 1669 – Newton used. We favor the latter – because its solar positions match Yahuda 24E better. Then 1669 must be the lower time bound for W.

It is our guess that Newton's "astronomical exercises" mentioned in Edward Sherburne's book in 1673 was the lunar visibility theory developed in N. If yes, then Newton certainly could have valued these 'exercises' highly if in 1673 he had thought of going public with a proposed appendix to the coming book by Nicholas Mercator – then a celebrity among British natural philosophers. The fact that Newton did not bother to recompute the table in N from the Flamsteed-Horroxian 1672 tables published by Wallis a year later suggests 1673 as a plausible upper time bound for N.

---

[65] J.P. Pratt. "Newton's Date for the Crucifixion." *Q. J. R. Astr. Soc.* 32 (Sept 1991), 301-4.

A remarkable fact is that the date for the Passion proposed in N, April 23, 34 AC, had already been proposed by no less an authority than Scaliger! This is easily overlooked, because Scaliger argued for it only in the 1[st] edition of *De Emendatione Temporum* (Paris 1583, pp. 262-265; same pages in Frankfurt 1593-edition) while in the 2[nd] edition of 1598 he changed his mind to April 3, 33 AC.[66] Newton owned the 1[st] edition which is dog-eared and annotated by him.[67] Since both editions of Scaliger's book were also owned by Isaac Barrow,[68] Newton did not need to buy it for himself until Barrow's death in May 1677. In case Newton borrowed his idea from Scaliger, he still has been working on N after 1677.

Indeed, the work on the Passion, it seems, did not stop in 1673, but spurred Newton's activity in the same direction. Two pages from Newton's notebook, known as Keynes MS 2, written in English and entitled "Christi Passio, Descensus, et Resurrectio," are dated to 1675.[69] Their text, however, deals with general Christian theology and not the chronology of the Passion, as does Yahuda 24E.

The fact that Cambridge does not own a single copy of the 1669 (Paris) edition of de Veil's Latin translation of Maimonides' *Sanctification of the New Moon* forced us to seek to examine Barrow's library. In vain – Barrow did not own any of Maimonides' books. An interesting conjecture would be to claim that Wickins owned a 1669 copy of Maimonides, but there is little chance to verify it. If, however, Newton consulted only the 1683 (London) edition of de Veil's translation of Maimonides, then 1683/4 must be an upper time bound for N. We have no hint that Newton dealt with this problem after 1684, when he gradually came into the limelight.

Of course, it would be interesting to connect N with Newton's mental crisis of 1693. At that time he seems to "clean out" his theological in-boxes. In 1690, he made an abortive attempt through Locke to publish in Amsterdam a theological tractate, *Two Notorious Corruptions of Scriptures*, which he chose to withdraw two years later. Besides, Newton is known to have composed in 1694 his "Origins of Gentile Theology."[70] To suspend the authority of St John could have been a hard blow to his religious creed. Since Newton did not change his astronomical table, copying it from W, the table most likely came from Wing's 1669 book, since it more closely approximates Kepler's area law, a pillar of Newton's theory in *Principia*, rather than Wing's earlier, 1651 book.

---

[66] Philipp Nothaft, personal communication of July 25, 2008. In the 2[nd] edition Scaliger corrected an error made in the 1[st] one, where he had placed the epoch of the Jewish calendar (BaHaRaD) a year later. The latter might shrink in disrepute and oblivion in the eyes of the community of scholars. Not so with Newton's eyes!
[67] *The Library of Isaac Newton*, 233 (cat. No. 1454).
[68] See the Catalogue of Barrow's books in M. Feingold (ed.), *Before Newton* (Cambridge 1990), 341-70.
[69] See ref. in Westfall, 311. We thank John T. Young for making these pages available for us.
[70] See Westfall, 490-91 and 504.

## Conjectures

To cope with the reversed order of the 'fair copy' W and the 'rough draft' N, as well as the number '3' for the page on the top of W, we conjecture that there must exist an earlier (yet undiscovered) draft of the 'fair copy' buried somewhere among his alchemical manuscripts. In support, we point out the fact[71] that after buying manuscripts at Sotheby's 1936 auction, John Maynard Keynes and Avraham Sholom Yahuda met several times to exchange different lots of documents, which assumes a mixture of purely alchemical (Keynes) and purely theological (Yahuda) sets. Some confusion inevitably must remain.

There is also another, somewhat Borgesian conjecture. The 'fair copy' W was a draft written (and not just copied) by Wickins. It was he who read Lange. It was he who conceived the idea of checking the moon's visibility in the years around Jesus' death. However, he was unable to compute this on his own and asked Newton to do "some boring astronomical computations." This would explain the conflict between them c. 1678, which could have been of the same nature as that between Newton and Hooke eight years later:

> Now is not this very fine? Mathematicians that find out, settle and do all the business must content themselves with being nothing but dry calculators and drudges; while another that does nothing but pretend and grasp at all the things must carry away all the invention as well as those that were to follow him as of those that went before.[72]

In the same letter, facing serious conflict with Hooke about priority, Newton even threatened to withdraw the last part of the *Principia* from publishing. This story, which describes Newton's character best, might explain why Newton never published his conclusions from Yahuda 24E and never mentioned them later, in his *Observations*. Unfortunately, we know too little about Wickins' own talents and work to substantiate this conjecture.[73] Whether their conflict also had a theological layer – the priority of St. John over the synoptic gospels – we can only put forward as a possibility.

---

[71] Related by Richard Popkin in "Newton and Maimonides" (note 8).
[72] In a 1686 letter to E. Halley, see Westfall, 448.
[73] See Westfall, 342-3, for some meager information about Wickins.

# Epilogue

There is a well-known story that in refusing to accept Newton's manuscripts from Abraham Shalom Yahuda's collection for Harvard in 1942, George Sarton not only mentioned the impropriety of such a purchase in war-time, as "intellectuals have more important things to do than browse through Newton's silly religious writings," but he also remarked: "Newton's religious views were as irrelevant to his scientific views as Maimonides' medical opinions were to his rabbinical views."[74] Maimonides' name can enter into the discussion because numerous theological documents from the Yahuda's collection specifically refer to Maimonides, and Yahuda likely could have used it as a major argument for the importance of his collection to understanding Newton's work.

Though we are not yet aware of any reference to Maimonides' medicine in Newton's alchemy manuscripts, Newton's close acquaintance with Maimonides' first lunar visibility theory seems to be firmly proved here. It is remarkable that Newton was able to reformulate this difficult problem into more practical terms and give it a serious application, building upon an initial idea by Lange, thus becoming another (after Lange, Descartes etc) "applied mathematician" of his Age. The moon soon would occupy a central place in his thoughts, at least in 1694-1695, when he has been working on his own *Theory of the Moon*, a failure by his own account.[75] This could have been an additional reason that by the end of his life, Newton despaired of solving the Moon theory, seeing clearly the limits of his methodology in detecting the first lunar visibility. Fearing that it would become subject to pointed criticism, he totally suppressed all astronomical arguments in his *Observations* in favor of "common sense" arguments.

Despite Sarton's irony over Yahuda's collection, Yahuda MS 24E alone turns up a new layer of Newton's life, worthy of a chapter in a book. It connects several dispersed facts of Newton's life into a kind of detective story, where the protagonist is obsessed with finding the truth. It may not be complete: further study of Newton's handwriting in the margins of the Wing's 1669 book might not only confirm Whiteside's discoveries of Newton's early thoughts on planetary motions, but also provide a golden key to the precise date of his writing Yahuda 24E. Newton's dependence on Scaliger must be studied further.

---

[74]  See Popkin, "Newton and Maimonides" (note 8).
[75]  In that work, published by David Gregory in 1702, Newton never achieved the desired precision of 2', but only of 6' (according to Halley) or even 10' (according to Flamsteed). See Westfall, 547-8.

**Acknowledgments and Remarks**

The authors are grateful to Ayval Leshem (Bar-Ilan University); Joan Griffith (Annapolis, Md.); Robert van Gent (University of Utrecht); Peter Nockolds (London); A.R. (Tom) Peters (Amsterdam); Heiner Lichtenberg (Bonn); John T. Young (Newton's Project); Stephen Snobelen (University of King's College, Halifax), Alexander Gordin (Bar-Ilan University), Yaqov Loewinger and Maghdi Shemuel (both Tel Aviv); Michael Gorodetsky (Moscow State University); Roger Friedman (Rare Book Studio, member: ABAA, ILAB); Curtis Wilson (St. John's College, Annapolis, Md.); Owen Gingerich (Harvard); Alex Gindes (N.Y.); Binyamin Krausz (Rare Books Department, Hebrew University, Jerusalem); Dennis Duke (Florida State University); Lars Gislen (University of Lund, Sweden); Philipp Nothaft (University of Munich); Consuelo W. Dutschke and Jennifer Lee (Butler Library, Columbia University N.Y.).

# Appendix 1: Yahuda 24 E[1]

## Regulae pro determinatione Paschae

1. Mensis ille Nisan erat cujus dies 14 aequinoxium vernum proxime sequebatur.
Mensis cujusque proximus erat dies in cujus antecedente nocte Luna post
conjunctionem primo visa est vel sereno caelo juxta computum certo viderit potuit.
Si ad occasum solis altitudo lunae [supra] horizontem minor erat
quam 10 gr ea non potuit videri illa nocte, sin major quam 14
2. Judaei novilunium vernum definiebant ex Luna visa
[vel] compertis diebus 29 mensis, luna nocte subsequente visa
dies tricesimus constitutus est
.................................................................................................
3. Si ad occasum solis altitudo luna supra horizontem
in Judaea minor erat quam 10 gr vel potius 12 gr non potuit videri illa nocte. Sin major erat altitudo quam
14 gr potuit ea caelo sereno videri sed ob nubes incertam erat ejus
visio. Erit igitur aut nox illa ubi sole occidente altitudo Lunae supra horizon
tem proxime excessit 10 gr aut nox subsequens vesper primi Nisan. A nocte [tertia]
mensis non potuit incipere nam dies tricesimus primus nunquam adjectus est mensi praecedenti.
4. Sole arietem ingrediente si excessus long. Lunae supra long. Solis, vel minuatur
octogesima sui parte et reliquum augeatur sexta parte latitudinis borealis vel minu-
atur sexta parte latitudinis australis, colligatur altitudo lunae supra
horizontem Judaeae ad occasum solis quam proxime.
5. In primo Azymorum seu 15 die Nisan vesperi Dominus comedit Pascha cum
discipulis feria quinta Romana et eodem die judaico currente crucifixus est feria sexta
6. Ad horam tertiam po[st] meridianam Londini id est ad horam sextam vespertinam (seu duode-
cimam Judaicam) Hierosolymis loca solis et lunae annis AD 30, 31, 32, 33, 34 35 et
36 erant ut sequitur

| AC Martj | | Long Solis | Long Lunae | Dist L. a S. | Lat Lunae | Sole occ. alt | Vesper primi Nisan | Vesper Pascha seu 15 Nisan | |
|---|---|---|---|---|---|---|---|---|---|
| 30 | 23 | 0. 0. 45 | 0.10.33 | 9. 48 | 4.0 Austr | 9.48 | Mar 24, 25 | Apr 7,8 | fer 6,7 |
| 31 | 12 | 11.19.51 | 11.28.15 | 8. 24 | 3.35 Aust | 7.40 | Mar 13,14 | Mar 27,28 | fer 3,4 |
| 32 | 30 | 0. 8. 6 | 0.18.22 | 10. 16 | 0.33 Aust | 10.3 | Mar 31 | Apr 14 | fer 2 |
| 33 | 20 | 11.28.10 | 0.15.4 | 16. 54 | 0.49 Bor | 16.50 | Mar 20, 21 | Apr 3,4 | fer 6,7 |
| 34 | 9 | 11.17.10 | 11.24.15 | 7. 5 | 0.44 Bor | 7. 6 | Mar 10,11 | Mar 24,25 | fer 4,5 |
| 35 | 28 | 0.5.28 | 0.12.28 | 7. 0 | 3.39 Bor | 7.30 | Mar 29 | Apr 12,13 | fer 3,4 |
| 36 | 17 | 11.25.30 | 0.7.40 | 12. 10 | 4.21 Bor | 12.34 | Mar 17, 18 | Mar 31,32 | fer 7,1 |
| 34 Apr 8 Hor. 18.16' | | 0.16.20 | 1.3.10 | 16. 50 | 3.44 Bor | 17.15 | Apr 8, 9 | Apr 22,23 | fer 5,6 |

His excluduntur anni omnes praeter annum 34. Nam anno 30, Martij die 23 non potuit luna videri
nisi velis ipsam unico tantum die delituisse, et hora decima oct. post veram
conjunctionem se aparuisse idque in magna australi latitudine constitutam. Quam casum puto
neutiquam admittendum esse. De caeteris annis non est locus disputationis
nisi velis ipsorum inito ob fruges nondum maturescentes mensem intercalari qua ratione dies 15 Nisan incidere potuit in feriam Romanam an-
nis solis 31 34 et 35. Sed annus 35 sero incepit ut si
mensis intercalaris ei praeponetur messis fere praeteriret ante mensem Nisan. Sola est
ambiguitas in annis 31 et 34. Et praeferendus est annus 34 quoniam is
tam sine intercalatione extraordinaria quam cum intercalatione
respondet tempori praeposito. Cum intercalatione vero annus iste ita se habet
A.D. 34 Apr 8 hora Romana post meridiem Judaicam 6.16 longit S 0.16gr.20'
Long L 1.3.10 Dist L a S 16.50 Lat. L 3.44 Bor. Altitudo L supra horiz.
ad occasum solis 17.15. Ergo Vesper primi Nisan Apr 8 et Vesper 15 Nisan
Apr 22 fer 5

(p. 26)
Judaei novilunium mensis Nisan definiebant ex luna visa. Si completis diebus
29 Mensis Adar Luna ea nocte visa fuit dies tricesimus constitutus est primus dies
Mensis Nisan: sin Luna illa nocte visa non fuit, adjectus est mensi Adar dies ille
tricesimus et dies tricesimus primus constitutus est primus dies Nisan.

A solstitio brumali ad aestivum Luna nunquam videri potuit nisi ubi excessus longitudinis
ejus et longitudinis solis occidentis major erat quam 9 gr, expectanda erat ejus visio caelo sereno
ubi excessus ille ad occasum solis major erat quam 15 gr. Differentia
inter 9 et 15 oriebatur e latitudine Luna. Ubi sol erat in signis Capricorni et Geminorum
[in]itio proximae lunae in latitudine
australe maxima constituta videri non potuit nisi excessu longitudinis majori
quam 14½ gr, in maxima boreali videri non potuit nisi excessu longitudinis majori quam 9 gr

At sole in aequinoctio verno constituto minor erat ratio latitudinis. Nam
luna tunc in latitudine maxima australi videri non potuit nisi excessu longitudinis majori quam
11 gr, in latitudine vero max. boreali videri non potuit nisi excessu longitudinis majori
quam 9½ gr. Quod si sol versaretur
in aequinoctio autumnali, Luna in maxima lat. australi videri non potest
nisi excessu longitudinis 24 gr aut majori, in maxima vero latitudine boreali
videri non potest nisi excessu longitudinis majori quam 10 gr.
Quod si excessus longitudinis lunaris supra solarem quovis tempore
major esset quam qui hic assignatur, expectanda erat apparitio Lunae eadem nocte.

Atque haec ita esse intelligunt astronomiae periti qui cum rationibus astronomi
is conferent ea quae Majmonides lib. de ratione Intercalandi cap. 17 sect 3 et 4 scripsit.

Dependet enim apparitio Lunae ab ejus altitudine supra horizontem ad occasum solis
[ita] ut ubi altitudine Lunae ad occasum centri solis, non
major sit quam 10¼ gr apparitio ejus illa nocte ex-
pectanda non sit, ubi vero major est quam 10¼ gr ea expectanda sit in locis montanis per
omnem Judaeam. Generaliter tradit Majmonides Lunam in Judaea nunquam apparuisse nisi
ubi ad horam visionis major erat differentia longitudinum Solis et
Lunae quam 9 gr. Sit eo tempore borealis latitudo lunae 5 gr. Et distantia lunae a sole
erit 10gr 18'. Immineat luna soli perpendiculariter (quod fieri potest in signis Geminorum et
Capricorni) et altitudo ejus supra
horizontem erit 10gr 18' nec tamen videri potest illa nocte. Quod si
Sol ..... a solstitio aestivo ad brumale[m] in illis sex signis Luna, ait
Majmonides, non videbitur nisi differentia longitudinum sit plus quam 10 gr.
Sit ea 10 gr et latitudo lunae borealis 5 gr et altitudo lunae supra horizontem
ubi sol solsti[ti]o alterutri proximus est erit 11 gr 6'. Nec tamen luna videri potest. Unde merito limitem altitudinis statuamus 11 gr vel ad
minimum 10½ gr   Potest tamen limes illae [gradu] uno minor esse ubi sol versatur in aequi-
noctio autumnali et Luna est in latitudine maxima australi [idque] propter magnam distantiam
Lunae a Sole
................................
[Adhibitus] igitur 10½ gr requiritur ad visionem lunae
ea nocte ubi distantia Lunae a sole non major est quam 12 gr.

Jam vero ad occasum solis altitudo Lunae accurate satis inveniatur si differentia
longitudinum minuamur octogesima sui parte et reliquum dein vel minuatur sexta
parte latitudinis australe vel augeatur sexta parte latitudinis borealis.

# English Translation

p. 25

## RULES FOR THE DETERMINATION OF EASTER

1. The month Nisan was the one whose 14th day followed most closely the vernal equinox. If at sunset the altitude of the Moon over the horizon was less than 10°, it could not be seen that night, but if it was greater than 14°...

2. The Jews determined the spring New Moon by its sighting …
[[*several lines follow with many amendments of difficult interpretation*]]

3. If at sunset the Moon's altitude over the horizon in Judea was less than 10°, or rather 12°, it could not be seen that night. But if its altitude was greater than 14°, it could be seen if skies were clear, but if there were clouds, its visibility would be uncertain. Therefore, the eve of Nisan 1 would be either the night when the Moon's altitude over the horizon at sunset nearly exceeded 10° or the following night. The month could not start from the third night because there never can be a 31st day in the preceding month.

4. When the Sun enters Aries, if the excess of the Moon's longitude over that of the Sun is diminished in its eightieth part, and the result either increased by the sixth part of its northern latitude, or decreased by the sixth part of its southern latitude, we arrive at the Moon's approximate altitude over the horizon of Judea at sunset.[2]

5. On the first day of unleavened bread, also the eve of Nisan 15, the Lord ate the Passover with his disciples, the Roman's Thursday, and, as the same Jewish day advanced, he was crucified on Friday.

6. At the third hour after midday in London, which is the sixth hour in the afternoon (or the Jewish twelfth hour[3]) in Jerusalem, the places of the Sun and the Moon for the years 30, 31, 32, 33, 34, 35 and 36 were as follows:

| AC March | | Long. Sun | Long. Moon | Dist. M from S. | Lat. Moon | Alt. Moon at | Eve of Nisan 1 | Passover eve or Nisan 15 | |
|---|---|---|---|---|---|---|---|---|---|
| 30 | 23 | 0. 0. 45 | 0.10.33 | 9. 48 | 4.0 S | 9.48 | Mar 24, 25 | Apr 7,8 | Fr,Sa |
| 31 | 12 | 11.19.51 | 11.28.15 | 8. 24 | 3.35 S | 7. 40 | Mar 13,14 | Mar 27,28 | Tu,We |
| 32 | 30 | 0. 8. 6 | 0.18.22 | 10. 16 | 0.33 S | 10. 3 | Mar 31 | Apr 14 | Mo |

| AC March | | Long. Sun | Long. Moon | Dist. M from S. | Lat. Moon | Alt. Moon at | Eve of Nisan 1 | Passover eve or Nisan 15 |
|---|---|---|---|---|---|---|---|---|
| 33 | 20 | 11.28.10 | 0.15.4 | 16. 54 | 0.49 N | 16. 50 | Mar 20, 21 | Apr 3,4      Fr,Sa |
| 34 | 9 | 11.17.10 | 11.24.15 | 7. 5 | 0.44 N | 7. 6 | Mar 10,11 | Mar 24,25 We,Th |
| 35 | 28 | 0.5.28 | 0.12.28 | 7. 0 | 3.39 N | 7. 30 | Mar 29 | Apr 12,13 Tu,We |
| 36 | 17 | 11.25.30 | 0.7.40 | 12. 10 | 4.21 N | 12. 34 | Mar 17, 18 | Mar 31,32 Sa,Su |
| 34    Apr 8 18h 16m | | 0.16.20 | 1.3.10 | 16. 50 | 3.44 N | 17. 15 | Apr 8, 9 | Apr 22,23 Th,Fr |

From these, all years except year 34 must be excluded. Because in the year 30, on March 23 the Moon could not be seen, unless you think that only in that case the rule failed, and that the Moon became visible eighteen hours after the true conjunction, when it had a large Southern latitude. This case I think by no means can be accepted. Concerning the other years, there is no place for discussion, unless you want the start of those years to be delayed because of an intercalation due to the late ripening of the crops, in which case Nisan 15 could occur on [Friday] in the years 31, 34 and 35. But the year 35 started late, and if you admit an intercalation, the harvest would be ripe before the month Nisan. Therefore the only alternative is between the years 31 and 34. The year 34 must be preferred because without and with an extraordinary intercalation, it meets the conditions set before.

The conditions for this year with intercalation are as follows:
AD 34, April 8th, 6.16' Roman hours PM in Judea. Longitude of the Sun $0^s16°20'$. Longitude of the Moon $1^s3°10'$. Distance from the Sun to the Moon 16°50'. Latitude of the Moon 3°44' N. Altitude of the Moon over the horizon at sunset 17°15'. Therefore the eve of Nisan 1 was April 8, and the eve of Nisan 15 was Thursday April 22.

p. 26
The Jews determined the New Moon of the month Nisan by its sighting. If on the night following the 29th day of the month Adar the Moon was seen, then the 30th was established as the first day of the month Nisan; but if the Moon was not seen that night, that day was added to the month Adar and the thirty-first was established as the first of Nisan.

From the winter solstice to the summer solstice, the Moon could never be seen if the excess of its longitude over the longitude of the setting sun was not greater than 9°. With clear skies, visibility should be expected when the excess at sunset was greater than 15°. The distinction between 9° and 15° proceeded from the latitude of the Moon. When the Sun was in the signs Capricorn and Gemini, the Moon's appearance in its maximum southern latitude could only be seen if the excess in longitude was greater than 14½°; but if in its maximum northern latitude, it could only be seen if the excess in longitude was greater than 9°.

On the other hand, when the Sun was in the spring equinox, the influence of latitude was less, for then the Moon in its maximum southern latitude could not be seen unless the longitude excess was greater that 11°, while in its maximum northern latitude it could not be seen unless the excess was greater than 9½°.

And if the Sun was in the autumn equinox, then the Moon in its maximum southern latitude could not be seen unless the longitude excess was 24° or more, while in its maximum northern latitude it could not be seen unless the excess was greater than 10°. Thence if the excess of the longitude of the Moon over that of the Sun at any time was greater than the values assigned here, the Moon's visibility in that night was to be expected.

That this is so is understood by people skilled in astronomy, who adjust their astronomical computations to comport with the rules Maimonides wrote in his book 'On the Method of Intercalations', chapter 17, sections 3 and 4.[4]

The Moon's visibility, therefore, depends on its altitude over the horizon at sunset, so that when the altitude of the Moon at the setting of the Sun's center is not greater than 10¼°, its visibility is not to be expected; but if it is greater, it is expected to be seen from the high places all over Judea.

Maimonides says that in Judea, the Moon generally never appeared unless at the time of sighting the difference between the longitudes of the Sun and the Moon was greater than 9°. Let the latitude of the Moon at that time be 5° North; then the distance from the Sun to the Moon will be 10°18'. Let the Moon be placed perpendicularly over the Sun (which only can happen in the signs of Gemini and Capricorn); its altitude over the horizon will be 10°18' and then it will not be possible to see it that night.

If the Sun is in those six signs from the summer solstice to the winter solstice, the Moon, says Maimonides, will not be seen unless the difference in longitude is more than 10°. Let that difference be 10° and the Moon's latitude 5° North; then the Moon's altitude over the horizon, when the Sun is near any of the solstices, is 11°6', and the Moon is not visible. This is the justification for our limit of 11° or even 10½°. That limit could be one degree less when the Sun is near the autumn equinox and the Moon is in its maximum southern latitude, because of the great distance of the Sun to the Moon.

In summary, 10½° are required to see the Moon on a night when the distance from the Sun to the Moon is no greater than 12°.

Then the Moon's altitude at sunset can be found with enough accuracy, if the difference of the longitudes is diminished by one-eightieth and the result is either reduced by one-sixth of the southern latitude or incremented by one-sixth of the northern latitude.

(p. 27-29)

# Regulae pro determinatione Paschalis

1. Mensis cujusque primus erat dies in cujus antecedente nocte luna post conjunctionem primo apparuit vel [non?] sereno caelo juxta computum apparere potuit.

2. Si ad occasum solis Luna minus distabat a conjunctione quam 8 1/6 vel 9 grad, ea non potuit videri illa nocte, sin plusquam 15 vel forte 14 grad in consequentia distabat certo potuit videri. In intermediis distantiis incerta erat ejus apparitio nisi quatenus a latitudine et celeritate motus determinare licuit. Utpote cum maximam haberet Australem latitudinem, distantia ab occidente sole citra quam non potuit illa nocte videri erat 10 1/6 aut 11 grad circiter, et distantia ultra quam certo potuit videri erat 14 aut ad summum 15 grad praesertim si in Apogaeo existens tardissime movebatur. Cum vero maximam habebat Borealem latitudinem, lim[ites] illi erant 9 et 13 grad circiter aut forte 8 1/6 et 12 p[rae-] sertim si in Perigaeo existens celerrime movebatur.

in intermediis latitudinibus limites erant proportion[es]
intermedii. Vide citationes ex Elia et Majemonid[es]
pag: 126 et 127 Langii de Anni Christi

3. Mensis ille Nisan erat cujus dies 14 aequinoxium ver[num]
proxime sequebatur.

4. In primo Azymorum seu 14[15] die Nisan Vesperi Dominus
comedit Pascha cum discipulis feria quinta Roman
et eodem die judaico feria sexta Roman. passus est.

5. Synedrium sedulo definiebat neomenias per visiones
ut manifestum est missione Testium in Montana et
examinatione de Phasi Visa Langius pag 112, 113 et 116.
Et mutatione Neomeniae statutae quoties e remotis re-
gionibus adventantes testabantur se lunam vidisse
in 30 die superioris mensis quae incolis Hierusalem
non apparuit ante diem 31 pag. 399 Langii.

Juxta has regulas Tabula sequens componitur, in
quo loci solis et lunae ad occasum solis in horizonte
Hierusalem hoc est... horam tertiam P. M. Londini
circiter proxime post conjunctionem sub initio Mensis
Nisan Annis Domini 30 31 32 33 34 35 36 et 37 e[xhi-]
bentur: et pro distantia Lunae in consequentia [vespere]
primi Nisan assignatur una[m] cum feria Romana in
quam vesper diei 14 incidit.

| AD Martij | | Long: S | Long: L | Dist: L a S | Latitud: L | Vesper primi Nisan | Vesper Paschalis seu 14[15] Nisan |
|---|---|---|---|---|---|---|---|
| | | s  gr  m | s  gr  m | gr  m | gr  m | | |
| 30 | 23 | 0. 0. 45 | 0.10.33 | 9. 48 | 4. 0 Austr | Mart: 24 | Apr: 6 ,      feria 5[6] |
| 31 | 12 | 11.19.51 | 11.28.15 | 8. 24 | 3.35 Austr | Mart: 13 | Mart: 26[27], fer. 2[3] |
| 32 | 30 | 0. 8. 6 | 0.18.22 | 10. 16 | 0.33 Austr | Mart: 30 [vel] 31 | Apr: 12[13]  f. 7[1] vel 13[[14]], f[1[2]] |
| 33 | 19 | 11.27.11 | 0. 0. 6 | 2. 55 | 0.26 Austr | * | * |
| 33 | 20 | 11.28.10 | 0.15. 4 | 16. 54 | 0.49 Bor | Mart: 20 | Apr: 2[3], fer 5[6] |
| 34 | 9 | 11.17.10 | 11.24.15 | 7. 5 | 0.44 Bor | Mart: 10 | Mart: 23[24], fer 3[4] |
| 35 | 28 | 0.5.28 | 0.12.28 | 7. 0 | 3.39 Bor | Mart: 29 | Apr 11[12]}, fer 2[3] |
| 36 | 17 | 11.25.30 | 0.7.40 | 12. 10 | 4.21 Bor | Mart: 17 vel 18 | Mart: 30[31], fer 6[7] vel  31[32], fer 7[1] |
| 37 Apr 4 | | 0.12.46 | 0.14.56 | 2. 16 | 5. 0 Bor | Apr 5 | Apr: 18[19], fer 5[6] |

Juxta hanc tabulam Pascha non incidit in fer 5 Rom
nisi in annis 30, 33 et 37. Adeoque si anni 30 et 37 per Histo-
riam excludi possint solus restabit 33 in quo Dominum
pati liceret. Et in hoc anno Martij 19 ad occasum solis
Luna ab eo distabat 2 grad 55 min tantum adeoque illa nocte videri non potuit. At in sequente nocte Martij 20 distabat
quasi 17 grad, et proinde tunc certo potuit videri praesertim
cum ad Boream verserit latitudine 49 min. Id itaque erat vesper
primi Nisan, et inde 14 Nisan vespere, quando Judaei im-
molarunt Pascha incidit in Apr: 2 fer 5 [sty]lo Roman:
et post mediam noctem incidit in Apr: 3 fer: 6; (In) quo
itaque Dominum pati licuit.

E ceteris Annis Pascha Anni 36 proxime ac-
cessit ad fer 5, quippe quod in fer 6 juxta calculum
incidere potuit. Sed non potuit incidere in fer 5 quia
sic antecedens neomenia visibilis fuisset Martij 16,
cum tamen luminaria nondum fuerint conjuncta Lunaeque
defuerint 9 grad longitudinis minimum quo consp[icari] potuisset.

Exclusis itaque annis 30 et 37 solus restavit 33, nisi
forte annus passionis fuerit intercalaris. Videamus
itaque in quibus annis vesper 14 proximi mensis incideri po-
tuit in feriam quartam. Scilicet Neomenia in 29 vel
ad lunam 30 diebus semper redit, hoc est, una vel ad summum
duobus diebus post quatuor hebdomadas; et proinde
annis 31 et 35, cum vesper 14 prioris mensis juxta tabula[m]
inciderit in feriam 2, vesper ille sequentis Mensis incid[ere]
debuit in feriam 3 vel ad summum in feriam 4.   Sic ann[us]
32 [ubi] 14 prioris Mensis incidit in fer 7 vel 1 idem vesper
sequentis Mensis incidere debuit in fer 2. Et anno 36 in
14 prioris incidit in fer 6 vel 7 [idem] vesper sequentis M[ensis]
incidere debuit in fer 1. Sed de anno 34 [res] non ita ce[nseo?]
per hanc regulam determinatur. Ideoque calculum subjunxi

| AD Apr. | hor. | min. | Long. $S^5$ | | | Long. L | | | Dist. S a L | | Lat. L | |
|---|---|---|---|---|---|---|---|---|---|---|---|---|
| | | | s | gr | m | s | gr | m | gr | m | gr | m |
| 34 | 8 | 18 | 16 | 0.16.20 | | 1. | 3.10 | | 16 | 50 | 3. 44 Bor | |

Luna cum hic fuerit in maxima fere latitudine Borea[ale]
per Reg 2 certo potuisset videri si tantum grad 13 vel forte
12 distasset a sole praesertim cum jam fuerit in perigaeo
ut calculus indicat. Et multo magis [videri] necesse est cum
adhuc longius quatuor gradibus circiter distaverit, nempe 16 [grad]
50 min. Quam in hoc vespere Apr: 8 collocanda est neomen[ia]
et non in proxumum differenda. Adeoque vesper 14 hujus Me[nsis]
incidit in Apr: 21 quae est feria 4. Solus itaque restat annus
33 in quo vesper 14 Nisan (sive proxime sequebatur
AEquinoxium sive, Anno intercalato, delatus erat in [se-]
quenter lunam) incidit in feriam 5, ideoque in quo Dom[inum]
pati licuit.

De veritate tabularum Astronomicarum etsi ad tempor[a]
tam longe praeterita applicentur non potest esse dubi-
tatio, siquidem ecclipses et lunares appulsus ad st[ellas]
fixas sub idem fere tempus observatas tam bone de-terminant ut vix credibile sit ultra unum vel ad
sumum duos gradus errare posse.

pp. 27-29

# RULES FOR THE DETERMINATION OF EASTER

1. The first day of the month was the one following the night when the moon first appeared after the new moon conjunction or, if the sky was not[6] clear, when the computation showed that it could have been seen.

2. If at sunset the distance of the Moon from the Sun[7] was less than 8°10' or 9°, it could not be seen that night; if the distance was larger than 15° or perhaps 14°, surely it could be seen. In intermediate distances, its visibility was uncertain; insofar it could not be determined from the latitude and the celerity of its motion. When it was at maximum Southern latitude, the distance from the setting Sun less than which the Moon could not be seen was approximately 10°10' or 11°, and the distance over which it could be seen with certainty was 14°, or at most 15°, especially when moving more slowly in apogee.[8] When it reached maximum Northern latitude, however, those limits were approximately 9° and 13°, or perhaps 8°10' and 12°, especially when it moved faster in perigee.[9] In intermediate latitudes, the limits were proportionally intermediate. See quotations from Elia[10] and Maimonides in pages 126 and 127 of Lange's *De Annis Christi*.[11]

3. The month of Nisan was the earliest one in which its 14th day followed the spring equinox.

4. In the first day of unleavened bread, which was the eve of Nisan[12] 14,[15] the Lord ate the Passover with his disciples, a Thursday in the Roman reckoning; on the same day according to the Jews, a Friday according to the Romans, the Passion took place.

5. The Sanhedrin carefully defined the New Moons by actual sightings as is manifested by the sending of witnesses to the high places and by examination of the phases that were reported (Lange, pages 112, 113, and 116). Also, they would change the already established New Moons when someone coming from remote regions testified that he had seen the Moon on the 30th day of the previous month, though it had not been seen by the inhabitants of Jerusalem until the 31st (Lange, page 399).

In accordance with these rules, the following table has been compiled with the places of the Sun and the Moon at sunset on the horizon of Jerusalem, that is, approximately at 3 p.m. London time, closely after the conjunction at the beginning of the month Nisan for the years 30, 31, 32, 33, 34, 35, 36, and 37 AD Also, in accordance with the distances of the Moon, the first day of Nisan is assigned together with the day of the week in which the evening of Nisan 14 falls.

| AC | Mar. | Long. Sun[13] | | Long. Moon | | Dist. Sun to Moon | | Latitude Moon | | Nisan 1 eve | Passover eve[14] or Nisan 14[15] |
|---|---|---|---|---|---|---|---|---|---|---|---|
| | | s | gr | s | gr | gr | m | gr | m | | |
| 30 | 23 | 0. | 0.45 | 0.10.33 | | 9. | 48 | 4. | 0  S | Mar. 24 | Apr. 6[7]  Thu[Fri] |
| 31 | 12 | 11.19.51 | | 11.28.15 | | 8. | 24 | 3. | 35  S | Mar. 13 | Mar. 26[27]  Mon[Tue] |
| 32 | 30 | 0. 8. 6 | | 0.18.22 | | 10. | 16 | 0. | 33  S | Mar.30or31 | Apr.12[13] Sat[Sun]  Apr.13[14] |
| 33 | 19 | 11.27.11 | | 0. 0. 6 | | 2. | 55 | 0. | 26  S | * | * |
| 33 | 20 | 11.28.10 | | 0.15. 4 | | 16. | 54 | 0. | 49  N | Mar. 20 | Apr: 2[3]  Thu[Fri] |
| 34 | 9 | 11.17.10 | | 11.24.15 | | 7. | 5 | 0. | 44  N | Mar. 10 | Mar. 23[24]  Tue[Wen] |
| 35 | 28 | 0. 5.28 | | 0.12.28 | | 7. | 0 | 3. | 39  N | Mar. 29 | Apr. 11[12]  Mon[Tue] |

| AC Mar. | Long. Sun [13] | Long. Moon | Dist. Sun to Moon | Latitude Moon | Nisan 1 eve | Passover eve [14] or Nisan 14 [15] |
|---|---|---|---|---|---|---|
| | s    gr | s    gr | gr   m | gr   m | | |
| 36       17 | 11.25.30 | 0. 7.40 | 12. 10 | 4. 21  N | Mar.17or18 | Mar. 30[31] Fri[Sat]   Mar. 31[1] Sat[Sun] |
| 37 Apr 4 | 0.12.46 | 0.14.56 | 2. 16 | 5.0  N | Apr. 5 | Apr. 18[19]   Thu[Fri] |

According to this table, Passover does not fall on Thursday except in the years 30, 33, and 37. Insofar as we can exclude the years 30 and 37 for historical reasons, it leaves only year 33 in which the Lord was allowed to suffer. And in that year on March 19 at sunset, the Moon's distance to the Sun was 2°55' -- so that in that night it could not be seen. But the next night, on March 20, the distance was almost 17°, which makes it certain that it could be seen, especially considering that the Moon deviated 49' in latitude to the North. It follows that this day was the eve of Nisan 1, and thence the eve of Nisan 14, when the Jews sacrificed the Passover, fell on April 2, a Thursday, and after midnight it was Friday, April 3, in which, therefore, the Lord was allowed to suffer.

Of the other years, the Passover of 36 came close to occurring on Thursday, though in fact, according to the computation, it fell on Friday. But it was not possible for it to occur on Thursday, because in that case the previous New Moon should have been visible on March 16, when the two luminous bodies had not reached conjunction yet - at least 9° more in the longitude of the Moon being required to make it visible.[15]

Excluding years 30 and 37, it only leaves the year 33, unless by chance the Year of the Passion had an intercalation month. Let us see in which years the eve of the 14th day of the previous month could occur on Wednesday. Of course the New Moon always comes back on either day 29th or 30th of the month, that is, at most two days over four weeks. Therefore, the years 31 and 35, in which the eve of the 14th of the previous month occurred on Monday, according to the table, the same eve of the next month should fall on Tuesday or on Wednesday at the most. In the year 32, where the 14th of the previous month fell on Friday or Saturday, the eve of the same day of the next month should occur on Sunday. But I think that the year 34 cannot be determined by this rule. Therefore I computed the following:

| AD Apr. | hour    min | Long. Sun [16] | Long. Moon | Dist. Sun to Moon | Latitude Moon |
|---|---|---|---|---|---|
| | | s    gr | s    gr | gr    m | gr  m |
| 34       8 | 18 | 0.16.20 | 1.3.10 | 16 | 3. 44 Bor |

As the Moon was then almost at its maximum Northern latitude, by Rule 2 it certainly could be seen as long as its distance to the Sun were only 13° or perhaps even 12°, especially if it had just passed perigee as the calculation indicates. There can be no doubt about its visibility, considering that its elongation was larger that that by more than 4°, that is 16°50'. Thence the New Moon must be placed on the evening of April 8 and not deferred to the next day. In that way, the eve of the 14th of that month occurred on Wednesday, April 21. The only year left is 33, when the eve of Nisan 14 (which either followed closely the equinox or, being an intercalary year, was deferred to the next lunation) occurred on Thursday, which is when the Lord was allowed to suffer.

The accuracy of the astronomical tables cannot be doubted, even for times so remotely in the past. Since they are able to determine so well the eclipses and close-ups (occultations) of the Moon with some fixed stars, observed at about the same epoch, it is hardly credible that they may err more than one degree or two at the most.

NOTES:

---

[1] Latin transcript, including bracketed corrections, and English translation by Eduardo Vila Echagüe; originals in the National Library of Hebrew University, Jerusalem. Endnotes by Eduardo Vila Echagüe and Ari Belenkiy.

[2] The angle formed by the ecliptic and the vertical line to the horizon at the moment of sunset reaches its annual minimum at the time of the vernal equinox, when it equals to the difference between the latitude of Jerusalem (32° N), and the obliquity of the ecliptic (23°), roughly 9°. To obtain the Moon's altitude at sunset, the ecliptical difference in longitudes and the latitude of the Moon must be projected to the vertical, using the formula: altitude = longitude difference · cos(9°) + latitude · sin(9°).  In the manuscript, cos 9° is approximated by fraction (1 - 1/80), while sin 9° - by fraction 1/6.

[3] Jewish practice in the $1^{st}$ century AD was to divide the daytime into 12 hours (John 11:9). Mishna *Berachot* (1:1) divides night in 3 *watches*. Maimonides in *Hilkot Kidush haKodesh* (6:2) speaks about a division of each day and night portion into 12 hours.

[4] Better known as *Hilkot Kidush haKodesh* or *Sanctification of the New Moon*, a part of a greater work, Mishnei Torah, written in 1170-1178.

[5] *Longitude is expressed as a Zodiacal sign (0-11), degree (0-30), and minute (0-60).*

[6] The word *not* is not visible in the manuscript, but is required by the meaning.

[7] In the text, literally a bit ambiguous, "from the conjunction of the Moon with the Sun."

[8] The Moon's reduced visibility in apogee is not related to its speed but to its smaller apparent size.

[9] The enhanced visibility in perigee is caused by the Moon's larger apparent size.

[10] Elia(hu) Bashyatzi. אדרת אליהו   [*Aderet Eliyahu*]. Constantinople, c.1540. [Republished in 1870 in Odessa by L. Hitche.] Known to Lange from excerpts in John Selden, *De anno civili & calendario veteris ecclesiae seu Reipublicae Judaicae dissertatio*, London 1644.

[11] Villum Lange [Wilhelmus Langius, 1624-1682], *De Annis Christi libri duo. Primus varios variorum gentium annos et tempora exponit. Secundus, in duas divisus partes, priori epochas nobiliores, quæ Christi ... passionem antecedunt, posteriori ipsum Dominicæ Nativitatis et Passionis tempus demonstrate* (Maire, Lugduni Batavorum [= Leiden], 1649).

[12] The manuscript shows a smaller number *15* over the 14 (shown here as an italicized superscript).

[13] Longitudes are expressed as Zodiacal **s**ign (0-11), **d**egrees and **m**inutes.

[14] *Most dates and weekdays of this column have smaller figures over them in the manuscript (shown here as italiced superscripts)*

[15] On March 16, 36 AD at sunset, the distance between the mean Moon and the mean Sun was 7 degrees, which, according to the criteria of his predecessors, would have placed the Moon near its limit of visibility, especially considering its large Northern latitude. However, the true positions of both bodies, as Newton argues, show them almost in conjunction, and therefore the New Moon could be visible only the next day.

[16] Longitudes are expressed as Zodiacal **s**ign (0-11), **d**egrees and **m**inutes.